\documentclass[twocolumn,prd,nofootinbib,floatfix,superscriptaddress]{revtex4}        
\usepackage{graphicx}
\usepackage{epsfig}
\usepackage{bm}
\usepackage[T1]{fontenc}
\usepackage[latin9]{inputenc}
\usepackage{amssymb}
\usepackage{float}
\usepackage{amsmath}
\usepackage{dcolumn}
\usepackage{cancel}
\usepackage[colorlinks]{hyperref}
\usepackage[usenames,dvipsnames]{color}
\hypersetup{
	breaklinks=true,
	pdfstartview={FitH},    
	colorlinks=true,       
	linkcolor=blue,          
	citecolor=red,        
	filecolor=magenta,      
	urlcolor=blue,           
	anchorcolor=green,      
	linktocpage=true
}

\newcommand{\be}{\begin{equation}}
\newcommand{\ee}{\end{equation}}

\begin{document}

\title{Black Hole Thermodynamics in Harada's inspired theory of gravity: Stability, Phase Structure and Geometrothermodynamics}

	
\author{S. Ghaffari}
\email{sh.ghaffari@maragheh.ac.ir}
\affiliation{Research Institute for Astronomy and Astrophysics of Maragha (RIAAM), 
University of Maragheh, P.O. Box 55136-553, Maragheh, Iran}

\author{G. G. Luciano}
\email{giuseppegaetano.luciano@udl.cat}
\affiliation{Departamento de Qu\'{\i}mica, F\'{\i}sica y Ciencias Ambientales y del Suelo, Escuela Polit\'ecnica Superior -- Lleida, Universidad de Lleida, Av. Jaume II, 69, 25001 Lleida, Spain
}

\date{\today}
\begin{abstract}
In this paper, we investigate the thermodynamic properties of spherically symmetric, static black hole solutions within the framework of Conformal Killing Gravity (CKG). This is a modified theory of gravity that retains all solutions of General Relativity, while addressing some of its theoretical shortcomings and enriching gravitational phenomenology at large distances. We derive key thermodynamic quantities, including mass, temperature, heat capacity and Gibbs free energy, to examine the stability and phase structure of  extended Schwarzschild-AdS and charged AdS black holes. Furthermore, employing the formalism of geometrothermodynamics, we analyze the behavior of the thermodynamic curvature scalar to identify critical points and characterize phase transitions. Our results demonstrate that the parameter \( \lambda \), which quantifies deviations from Einstein's theory, plays a pivotal role in shaping the thermodynamic behavior, resulting in new stability conditions and distinct phase transition patterns compared to those predicted by standard General Relativity.
	
	
\end{abstract}

\maketitle

\section{Introduction}

In theoretical physics, considerable effort has been devoted to developing a unified theory of gravity that can coherently describe the fundamental interactions between spacetime and matter. While General Relativity (GR) has proven remarkably successful in explaining gravitational phenomena at local and solar system scales, it faces significant challenges when extended to larger structures, such as galactic dynamics and the accelerated expansion of the Universe~\cite{Clifton:2011jh,Bull}.

To reconcile these discrepancies, two major components have traditionally been introduced into the standard cosmological model: dark matter and dark energy~\cite{Peebles:2002gy,Bertone:2004pz}. Although these components 
have proven highly effective in accounting for diverse astrophysical observations, they remain empirically elusive. This has led to growing concerns regarding the fundamental consistency of our current understanding of gravity and the overall composition of the Universe~\cite{Feng:2010gw,Frieman:2008sn}. In response, these conceptual limitations have motivated the development of alternative theories of gravity~\cite{Capoz}, which seek to explain cosmological phenomena without invoking unknown or unobserved constituents (see also~\cite{Jusufi,DiVal} for recent work along this direction).

Among the various alternative theories of gravity, one notable recent proposal is Cotton gravity, introduced by Harada as an extension of GR~\cite{Harada,Harada2}. Despite being in the early phases of exploration and facing several critical assessments~\cite{Crit1, Crit2, Crit3, Crit4, Crit5, Altas:2024pil}, the theory appears particularly compelling from a theoretical standpoint, as it engages with several foundational challenges inherent to GR and many of its extensions. Specifically, Cotton gravity (i) determines the cosmological constant as an integration constant that emerges dynamically from the equations of motion, (ii) derives the conservation of the energy-momentum tensor, $\nabla_\mu T^\mu_\nu = 0$, directly from the field equations rather than assuming it a priori, and (iii) ensures that conformally flat spacetimes are not necessarily vacuum solutions, thereby enriching the vacuum structure of the theory. 

Building upon these theoretical foundations, Harada has recently proposed a novel extension of gravity~\cite{Harada2a, HaradaNEW}. 
In this scenario, the conventional geometric term represented by the Einstein tensor is replaced by a rank-3 object, while the source term, typically given by the energy-momentum tensor, is instead expressed in terms of its gradients. 
The resulting field equations take the form $H_{\alpha\mu\nu}=8\pi {T}_{\alpha\mu\nu}$, where
\begin{eqnarray}
H_{\alpha\mu\nu}&=&\nabla_{\alpha}\mathcal{R}_{\mu\nu}+\nabla_{\mu}\mathcal{R}_{\nu\alpha}+\nabla_{\nu}\mathcal{R}_{\alpha\mu}\nonumber\\[1mm]
&&-\frac{1}{3}\left(g_{\mu\nu}\partial_{\alpha}+g_{\nu\alpha}\partial_{\mu}+g_{\alpha\mu}\partial_{\nu}\right)\mathcal{R}\,,\\[2mm]
T_{\alpha\mu\nu}&=&\nabla_{\alpha}T_{\mu\nu}+\nabla_{\mu}T_{\nu\alpha}+\nabla_{\nu}T_{\alpha\mu}\nonumber\\[1mm]
&&-\frac{1}{6}\left(g_{\mu\nu}\partial_{\alpha}+g_{\nu\alpha}\partial_{\mu}+g_{\alpha\mu}\partial_{\nu}\right)T\,.
\end{eqnarray}
Here, $R_{\mu\nu}$ is the Ricci tensor and $T_{\mu\nu}$ denotes the usual energy-momentum tensor, with their respective traces given by $R$ and $T$. It is important to note that $H_{\alpha\mu\nu}$ is completely symmetric in the indices $\alpha$, $\mu$, and $\nu$, and satisfies the traceless condition $g^{\mu\nu}H_{\alpha\mu\nu} = 0$. Similarly, $T_{\alpha\mu\nu}$ is fully symmetric and obeys the identity $g^{\mu\nu}T_{\alpha\mu\nu} = 2\nabla_{\mu}T^\mu_\alpha$, which in turn leads to the conservation law $\nabla_{\mu}T^\mu_\nu = 0$.

Shortly after, a parametrization has been identified in which Harada's equations are equivalent to the Einstein field equations modified by the inclusion of a divergence-free conformal Killing tensor (CKT) \cite{Molinari2}, namely
\begin{eqnarray}
    &&R_{\mu\nu}-\frac{1}{2}Rg_{\mu\nu}=T_{\mu\nu}+K_{\mu\nu}\,,\\[2mm]
    \nonumber
&&\nabla_{\alpha}K_{\mu\nu}+\nabla_{\mu}K_{\alpha\nu}+\nabla_{\nu}K_{\alpha\mu}=\\[2mm]
&&\hspace{0.5cm}\frac{1}{6}\left(g_{\mu\nu}\nabla_\alpha K + g_{\alpha\nu}\nabla_\mu K + g_{\alpha\mu}\nabla_\nu K  \right).
\end{eqnarray}
For this reason, the theory has been termed Conformal Killing Gravity (CKG). This reformulation explicitly reveals Harada's extension of General Relativity (GR) through the inclusion of a conformal Killing tensor, which satisfies the condition $\nabla^\mu K_{\mu\nu} = 0$ and enters the field equations as an additional source term.
 
It has been shown that the solution space of this theory encompasses all solutions of GR. Beyond this, the theory also yields novel configurations not present in GR, including generalized forms of the Schwarzschild metric \cite{Molinari}. Related studies have further identified a variety of vacuum cosmological spacetimes, as well as wormhole and black hole (BH) solutions, thereby broadening the spectrum of physically relevant geometries~\cite{Clement}.

On the other hand, recent research has increasingly interpreted BHs as thermodynamic systems, applying the laws of thermodynamics to gain insight into their physical behavior~\cite{Bekenstein:1973ur,Bekenstein:1974ax,Hawking:1975vcx,Hawking:1976de,Bardeen:1973gs,Gibbons:1976ue}. This perspective originates from the seminal works of Bekenstein and Hawking~\cite{Bekenstein:1973ur,Bekenstein:1974ax,Hawking:1975vcx,Hawking:1976de}, which established a deep connection between BH properties - such as horizon temperature and entropy - and thermodynamic principles. In this framework, BH entropy is proportional to the area of the event horizon, and the first law of thermodynamics is satisfied, implying that the event horizon area never decreases.
In parallel, cosmological observations indicate that the Universe is undergoing an accelerated expansion~\cite{Riess,Perlmutter}. One of the most widely accepted explanations  involves the inclusion of a cosmological constant $\Lambda$ in Einstein's field equations, effectively acting as a form of vacuum energy.

The presence of $\Lambda$ leads to generalized BH solutions with non-flat asymptotics. In particular, for positive $\Lambda$, the resulting geometry corresponds to Schwarzschild-de Sitter BHs, while for negative $\Lambda$, it yields Schwarzschild-anti-de Sitter BHs. These solutions not only extend the standard Schwarzschild metric but also provide a natural framework for exploring thermodynamic properties in spacetimes with constant curvature.

Recent developments in BH thermodynamics have led to the formulation of an extended phase space, in which the cosmological constant is interpreted as a thermodynamic pressure and its conjugate quantity is identified as a thermodynamic volume~\cite{Mann,Kastor:2009wy,Cvetic:2010jb}.
This reinterpretation allows the first law of BH thermodynamics to be generalized to include a $VdP$ term, thereby establishing a closer analogy with conventional thermodynamic systems. Within this framework, BHs exhibit rich phase structures, including critical behavior analogous to that of Van der Waals fluids, particularly in the anti-de Sitter case (see also~\cite{Luciano:2023fyr,Luciano:2023bai,Ghaffari:2023vcw,Sekhmani:2024udl} for recent applications in non-extensive thermostatistics).

The realization that asymptotically anti-de Sitter BHs can be interpreted via a gauge/gravity duality, specifically as thermal states within a corresponding dual field theory, has spurred the development of fluid-like models for describing their microscopic degrees of freedom~\cite{Maldacena,Witten,Chamblin}. This dual perspective has strengthened the analogy between BH systems and those in condensed matter physics, leading to increased interest in studying BH thermodynamics using geometric approaches, commonly referred to as \textit{geometrothermodynamics}~\cite{Davies:1977bgr,Cai:1998ep,Quevedo:2007mj,Quevedo:2008ry,Sahay:2010tx,Wei:2015iwa,Dehyadegari:2016nkd,Wei:2019yvs,Wei:2019uqg}.

Within this framework, insights into the microscopic interactions among the constituents of BHs are typically extracted by analyzing their macroscopic thermodynamic geometry. In particular, the Weinhold~\cite{Wein1} and Ruppeiner~\cite{Rupp1,Rupp2} metrics offer useful tools: the sign of the corresponding scalar curvature is interpreted as an indicator of the nature of underlying interactions. A negative curvature is generally associated with attractive interactions, whereas a positive curvature indicates repulsion. A vanishing scalar curvature suggests the presence of idealized systems with no dominant interaction type. Along this line, numerous studies have explored the thermodynamic geometry of a wide range of BH spacetimes, including (2+1)-dimensional Banados-Teitelboim-Zanelli (BTZ) BHs~\cite{Cai:1998ep},  Reissner-Nordstr\"om, Kerr and Reissner-Nordstr\"om--AdS BHs~\cite{Shen:2005nu}. The results reveal a rich thermodynamic structure, with the scalar curvature exhibiting different signs depending on the BH configuration.

Motivated by the aforementioned considerations, the present work is devoted to the investigation of BH thermodynamics and geometrothermodynamics within the framework of CKG. In particular, we analyze the thermodynamic behavior of static, spherically symmetric BH solutions supported by this extended theory of gravity. It is shown that the modifications introduced by CKG induce non-trivial corrections to key thermodynamic quantities, including the Hawking temperature, heat capacity and free energy. 
While the analysis is carried out at a theoretical level, the results point toward distinctive features that could, in principle, serve as potential signatures of the theory within BH physics. In this sense, the framework may offer a promising arena for probing the phenomenological implications of CKG beyond the purely classical regime.

The structure of this paper is as follows. In the next section, we review the exact solutions corresponding to Schwarzschild and Reissner-Nordstr\"om BHs in the context of CKG. Section~\ref{Ther} is dedicated to the derivation and analysis of various thermodynamic quantities associated with these solutions, whose behavior is further illustrated through graphical representations.
In Section~\ref{Stab}, we investigate the thermodynamic stability by examining the specific heat capacity and free energy, respectively. Section~\ref{Geo} focuses on the geometrothermodynamic analysis of the system. Finally, conclusions and possible future directions are presented in Section~\ref{Conc}.
Throughout this work, we adopt natural units.

\section{Static Spherically Symmetric Spacetime in CKG}
Let us consider the general static, spherically symmetric metric given by
\begin{equation}
\label{metric2}
ds^2 = -f(r)\,dt^2 + \frac{dr^2}{f(r)} + r^2\,d\theta^2 + r^2 \sin^2\theta\,d\phi^2.
\end{equation}
An extension of the Schwarzschild solution in a conformally flat spacetime was obtained by solving the CKG field equation under the assumptions of staticity, spherical symmetry and vacuum. In this setting, such an approach yields the nontrivial solution \cite{Harada2a,Molinari}
\begin{equation}
\label{f1}
f(r)=1-\frac{2M}{r}-\frac{\Lambda}{3}r^2-\frac{\lambda}{5}r^4\,.
\end{equation}
Here, the first term \( \frac{M}{r} \) corresponds to the Schwarzschild solution, while the second one \( \frac{\Lambda r^2}{3} \) indicates the presence of a de Sitter component, where \( \Lambda \) denotes a cosmological constant arising as an integration constant. 
The last term \( \frac{\lambda r^4}{5} \) introduces a correction characteristic of the underlying modified gravity theory, which becomes dominant in the asymptotic regime \( r \to \infty \). Although there is currently no strong empirical evidence constraining either the magnitude or the sign of \( \lambda \), it is generally expected that the parameter remains sufficiently small in absolute value to ensure compatibility with GR at low energies - a regime that is recovered in the limit \( \lambda \to 0 \) or for sufficiently small \( r \).

More recently, BH solutions in the framework of CKG coupled to nonlinear electrodynamics and scalar fields have been extensively investigated, leading to the derivation of a generalized Schwarzschild-Reissner-Nordstr\"om-AdS solution~\cite{Maxwell,Tarciso,Molinari}. 
The resulting metric is in the form
\begin{equation}\label{f2}
f_q(r) = 1 - \frac{2M}{r} + \frac{q^2}{r^2} - \frac{\Lambda}{3}r^2 - \frac{\lambda}{5}r^4\,,
\end{equation}
where the subscript \( q \) denotes the presence of a nonzero charge.
For \( \lambda = 0 \), this expression reduces to the well-known Reissner-Nordstr\"om metric with a cosmological constant. 


In order to explore the thermodynamic properties of the geometry~\eqref{metric2}, it is useful to consider the extended phase space framework, in which the cosmological constant is interpreted as a thermodynamic pressure, given by
\begin{equation}
\label{P1}
p = -\frac{\Lambda}{8\pi}\,.
\end{equation}
Since this identification yields a positive pressure for negative values of $\Lambda$, we shall henceforth restrict our attention to an Anti-de Sitter (AdS) background ($\Lambda < 0$).

\section{Thermodynamic Analysis}
\label{Ther}

This section presents a thermodynamic analysis of Schwarzschild-AdS and charged-AdS black holes within the CKG framework. To this end, we adopt a thermodynamic interpretation of the black hole mass \( M \), consistent with the extended first law in the canonical ensemble (fixed charge). Accordingly, we consider \cite{Mann}
\begin{equation}
\label{firstlaw}
dM = T\,dS + p\,dV\,,
\end{equation}
where \( M \) is understood as the enthalpy of the BH conceived as a thermodynamic
system, \( T \) and \( S \) denote the Hawking temperature and entropy, and \( p \) the pressure associated with a cosmological constant-like term, with \( V \) its thermodynamic conjugate volume.




A fundamental aspect of the thermodynamic analysis involves the definition of BH horizon entropy. In accordance with the Bekenstein-Hawking holographic scaling relation, the entropy of a BH is given by~\cite{Bekenstein:1973ur,Hawking:1975vcx}
\begin{equation}
\label{S}
S = \frac{A}{4} = \pi r_h^2\,,
\end{equation}
where \( A = 4\pi r_h^2 \) denotes the surface area of the BH horizon, with \( r_h \) being the radius of the event horizon.
The latter is determined by the condition \( f(r_h) = 0 \), where \( f(r) \) is the blackening function that characterizes the BH geometry. 

Therefore, making use of Eqs.~(\ref{f1}) and (\ref{f2}) together with the horizon condition, the mass of the Schwarzschild-AdS and charged AdS BHs can be expressed, respectively, as functions of the horizon radius as follows
\begin{eqnarray}
\label{MS1eq}
M(r)&=&\frac{1}{2}r-\frac{\Lambda}{6}r^3-\frac{\lambda}{10}r^5,\\[2mm]
M_q(r)&=&\frac{1}{2}r+\frac{q^2}{2r}-\frac{\Lambda}{6}r^3-\frac{\lambda}{10}r^5\,,
\label{MR1Seq}
\end{eqnarray}
where, for notational simplicity, the horizon radius has been denoted by \( r \). These expressions play a fundamental role in the thermodynamic analysis developed in the subsequent sections.

\begin{figure}[t]
\begin{center}
\includegraphics[width=8cm]{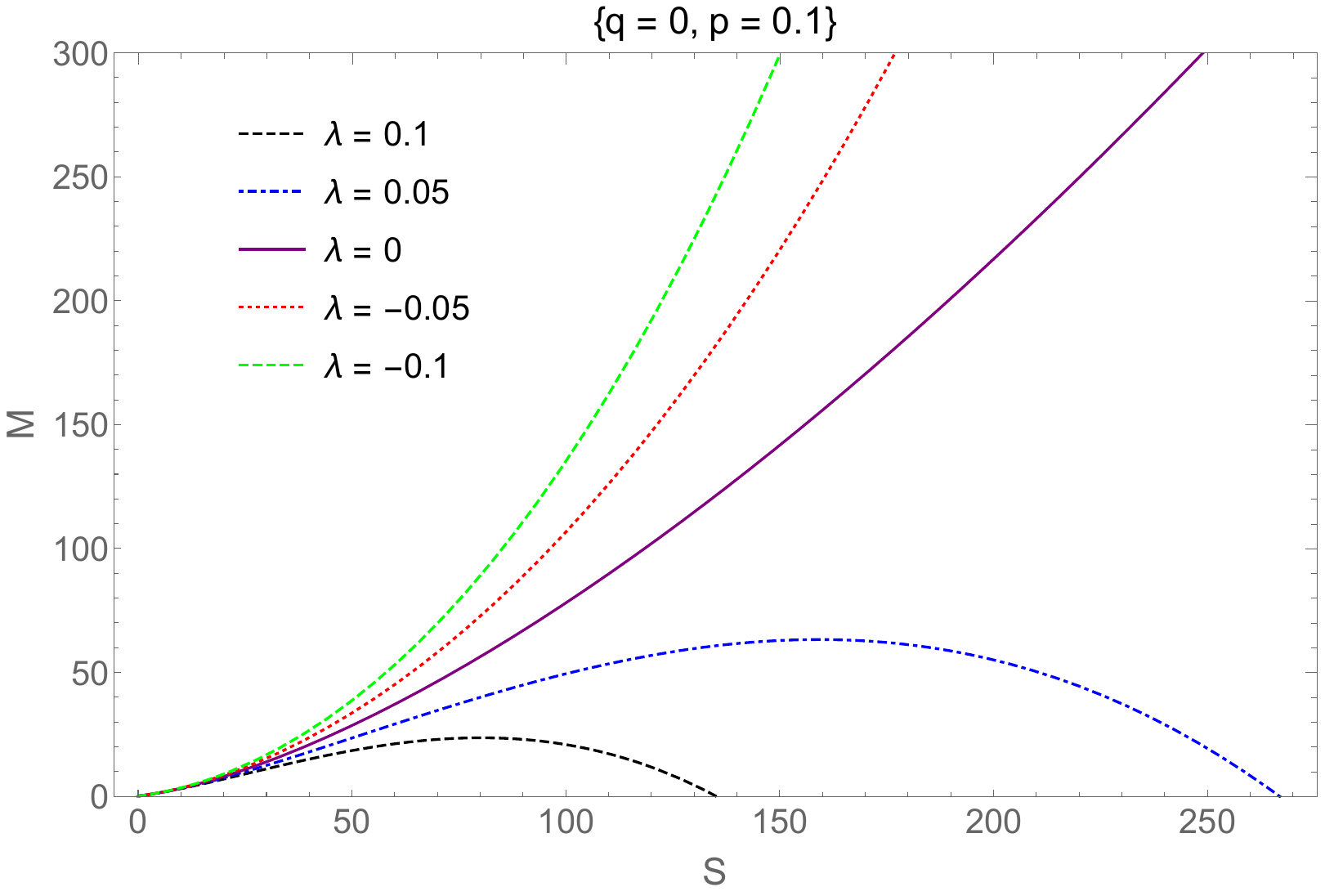} 
\caption{Mass of the Schwarzschild-AdS BH as a function of entropy, for various values of $\lambda$ and $p=0.1$.}
\label{figMS}
\end{center}
\end{figure}

\begin{figure}[t]
\begin{center}
\includegraphics[width=8cm]{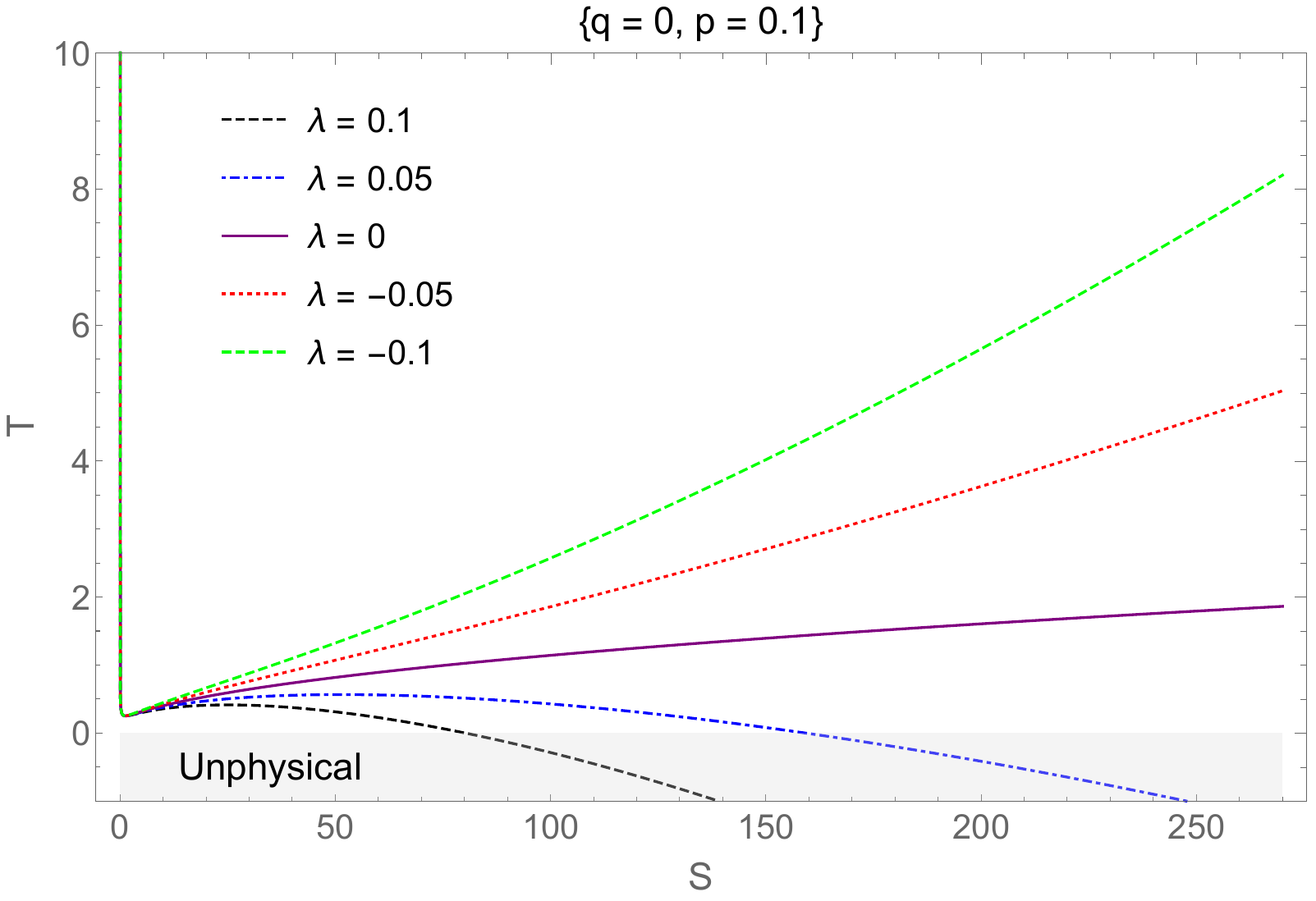}
\caption{Temperature of the Schwarzschild-AdS BH as a function of entropy, for various values of $\lambda$ and $p=0.1$. The shaded region corresponds to the unphysical regime where $T<0$.}
\label{figTS}
\end{center}
\end{figure}

\subsection{Schwarzschild-AdS black hole}

According to the Bekenstein--Hawking entropy-area relation~\eqref{S}, the Schwarzschild-AdS mass parameter given in Eq.~(\ref{MS1eq}) can be re-expressed in terms of the BH entropy as
\begin{equation}
\label{MS2eq}
M(S)=\frac{1}{2}\sqrt{\frac{S}{\pi}}\left(1+\frac{8pS}{3}-\frac{\lambda S^2}{5\pi^2}\right).
\end{equation}
The behavior of \( M \) as a function of entropy \( S \) is illustrated in Fig.~\ref{figMS}. Since our primary interest lies in exploring the effects of the modified gravity correction, we let the parameter \( \lambda \) vary while keeping the pressure fixed at \( p = 0.1 \). It is worth noting, however, that changes in \( p \) do not affect the overall qualitative behavior of the mass profile\footnote{We refer to~\cite{Mann,Luciano:2023fyr} for the specific values of the model parameters employed in this analysis.
}.

Two distinct cases can be identified. For \( \lambda \le 0 \), the mass increases monotonically with entropy, with the growth becoming faster as the absolute value of \( \lambda \) increases. This trend reflects the physical expectation that larger BHs (with larger horizon areas and thus larger entropy) require more energy to form, consistent with the interpretation of \( M \) as the enthalpy in extended BH thermodynamics. 

Conversely, for \( \lambda > 0 \), the system initially exhibits a regime in which the mass increases with entropy. However, beyond a certain BH size, the modified gravity correction becomes dominant, leading to a decrease in mass as entropy continues to grow, eventually driving it back to zero. Entropy values beyond this point would yield \( M < 0 \), which must be excluded on physical grounds.

This behavior represents a clear departure from standard BH thermodynamics. In particular, the fact that the mass decreases with increasing BH size suggests that the modified gravity correction may give rise to unconventional effects arising from higher-curvature or topological contributions, or may even indicate that such large BHs become unphysical within the classical regime of the theory (see the discussion below). This phenomenon could, in fact, signal the breakdown of the classical approximation itself, underscoring the necessity of quantum corrections or a more comprehensive theoretical framework to accurately describe the system beyond this regime.

To gain further insight into the thermodynamic properties of such systems, we now turn to the analysis of temperature. Examining the temperature behavior of BHs is crucial, as the slope and extrema of the \( T\text{-}S \) diagram provide information on local stability and signal the presence of possible phase transitions. Toward this end, by employing the first law of thermodynamics~\eqref{firstlaw}, the temperature is obtained by taking the partial derivative of the mass \( M \) with respect to the entropy \( S \), i.e.,
\begin{equation}
\label{TS1eq}
T(S)=\frac{\partial M}{\partial S}\Big|_p=\frac{\pi^2\left(1+8 pS\right)-\lambda S^2}{4\pi^\frac{5}{2}S^\frac{1}{2}}\,.
\end{equation}
The behavior of the temperature as a function of entropy is presented in Fig.~\ref{figTS}, plotted for various values of \( \lambda \) with fixed \( p = 0.1 \). Consistent with the evolution of the \( M\text{-}S \) profile, it is observed that for \( \lambda \leq 0 \), the temperature remains strictly positive. At low entropy, the temperature initially decreases, indicating a thermodynamically unstable phase (as the temperature drops while the BH absorbs energy). This trend continues until the system reaches a local minimum in temperature, which corresponds to a phase transition point. Beyond this minimum, the temperature increases monotonically with entropy, implying that the BH becomes thermodynamically stable in the large-entropy regime (as energy absorption now leads to a temperature increase).

A qualitatively similar behavior at relatively small entropy  is also observed in the case \( \lambda > 0 \). However, in this scenario, there exists a threshold value of \( S \) beyond which the temperature begins to decrease again, eventually reaching negative values (see the shaded region in Fig.~\ref{figTS}). This domain, corresponding to the regime where \( M \) becomes a decreasing function of \( S \), is clearly unphysical, as the absolute temperature, by definition, must remain strictly positive, \( T > 0 \).




\subsection{Charged AdS black hole}
\begin{figure}[t]
\begin{center}
\includegraphics[width=8cm]{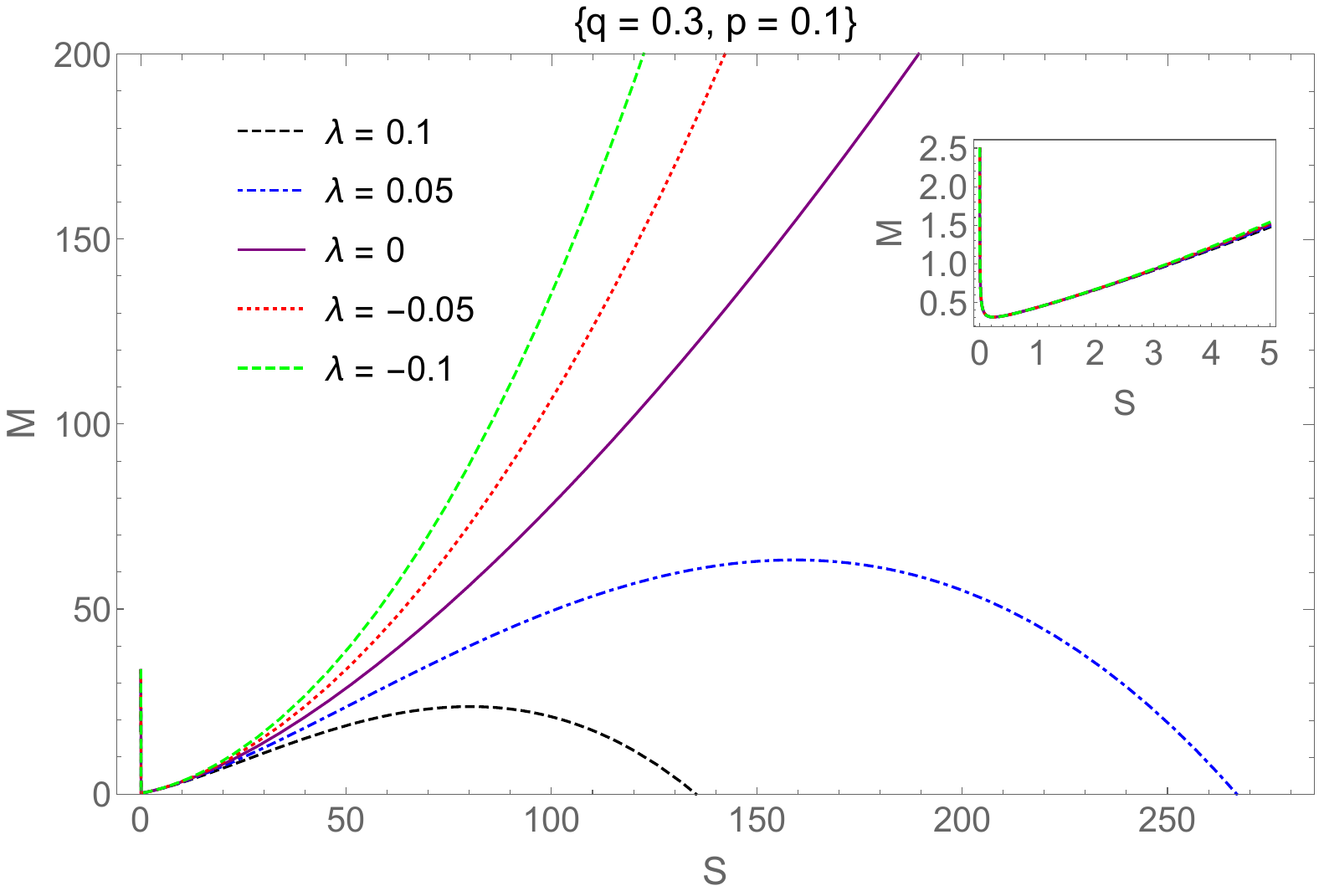}
\vspace{1mm}
\includegraphics[width=8cm]{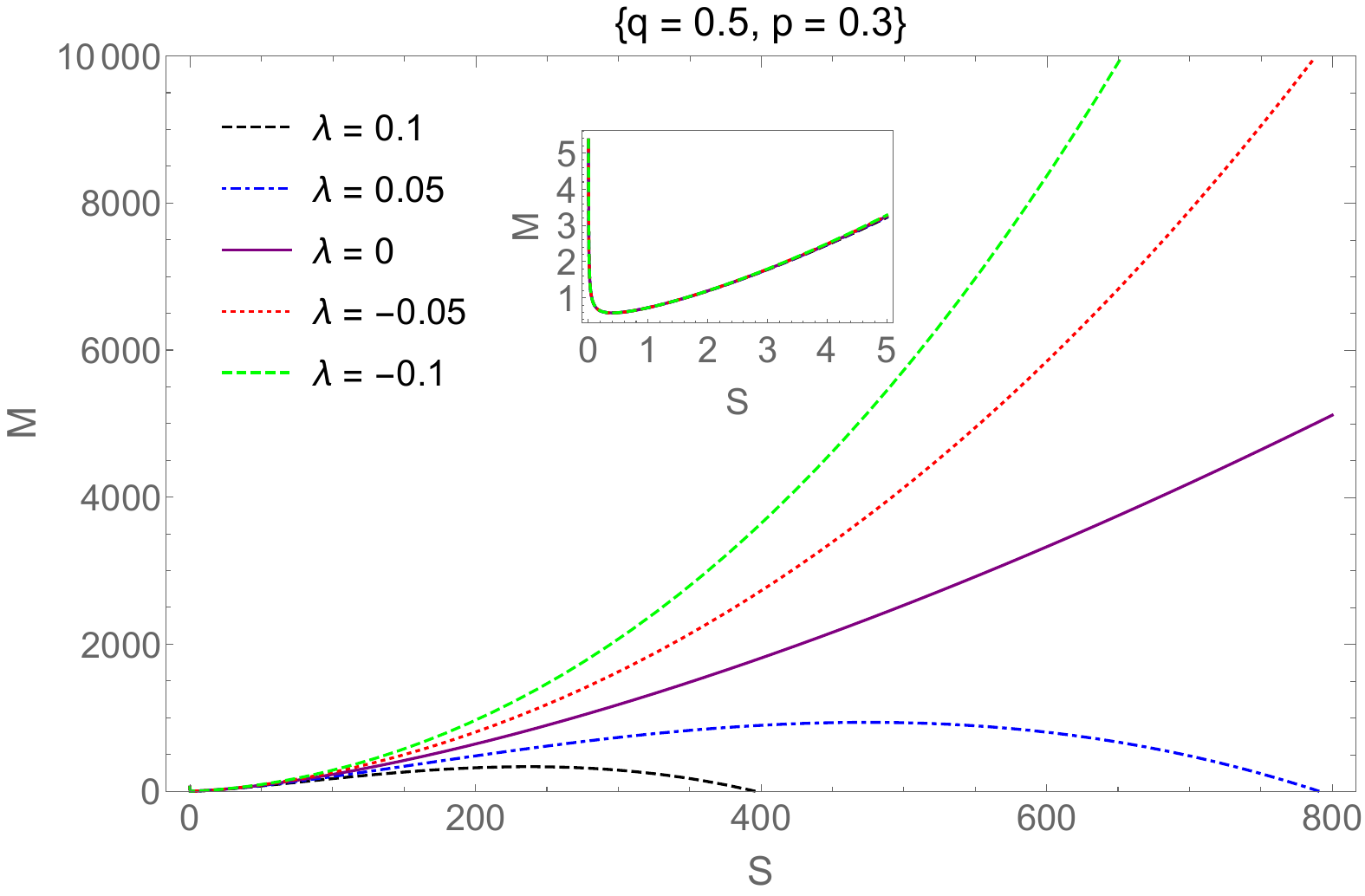}
\caption{Mass of the charged AdS BH as a function of entropy, for various values of $\lambda$, and $(q,p)=(0.3, 0.1)$ (upper panel) and $(q,p)=(0.5, 0.3)$ (lower panel).}
 \label{figMRN1}
\end{center}
\end{figure}

For charged AdS BHs, the mass expression given in Eq.~(\ref{MR1Seq}) can be reformulated in terms of the entropy as
\begin{equation}
\label{MR2eq}
M_q(S)=\frac{1}{2}\sqrt{\frac{S}{\pi}}\left(1+\frac{\pi q^2}{S}+\frac{8pS}{3}-\frac{\lambda S^2}{5\pi^2}\right).
\end{equation}
Figure~\ref{figMRN1} illustrates the behavior of the mass parameter as a function of entropy, for various values of \( \lambda \), \( p \) and \( q \). As shown in both panels, the BH mass initially decreases with increasing entropy\footnote{For charged AdS BHs, the $r \to 0$ limit leads to a divergent mass $M(r) \sim \frac{q^2}{r}$, as the charge contribution dominates at small scales, revealing a subtle behavior of the classical solution in this regime~\cite{Misner}.}, reaching a minimum before rising again. This represents a non-trivial difference with the behavior observed in the uncharged configuration examined earlier.
For the sake of comparison with recent literature, we emphasize that similar results have been reported in the context of quantum gravity-induced deformations of Boltzmann-Gibbs entropy~\cite{Fatima}, as well as in studies involving non-linear electrodynamics and Einstein-massive gravity~\cite{Nam}. 

Once again, for large BH sizes, the mass exhibits a monotonically increasing behavior for \( \lambda \leq 0 \), while for \( \lambda > 0 \) it begins to decrease beyond a certain entropy threshold. As previously discussed, this regime is unphysical, as it corresponds to negative values of the absolute temperature.

\begin{figure}[t]
\begin{center}
\includegraphics[width=8cm]{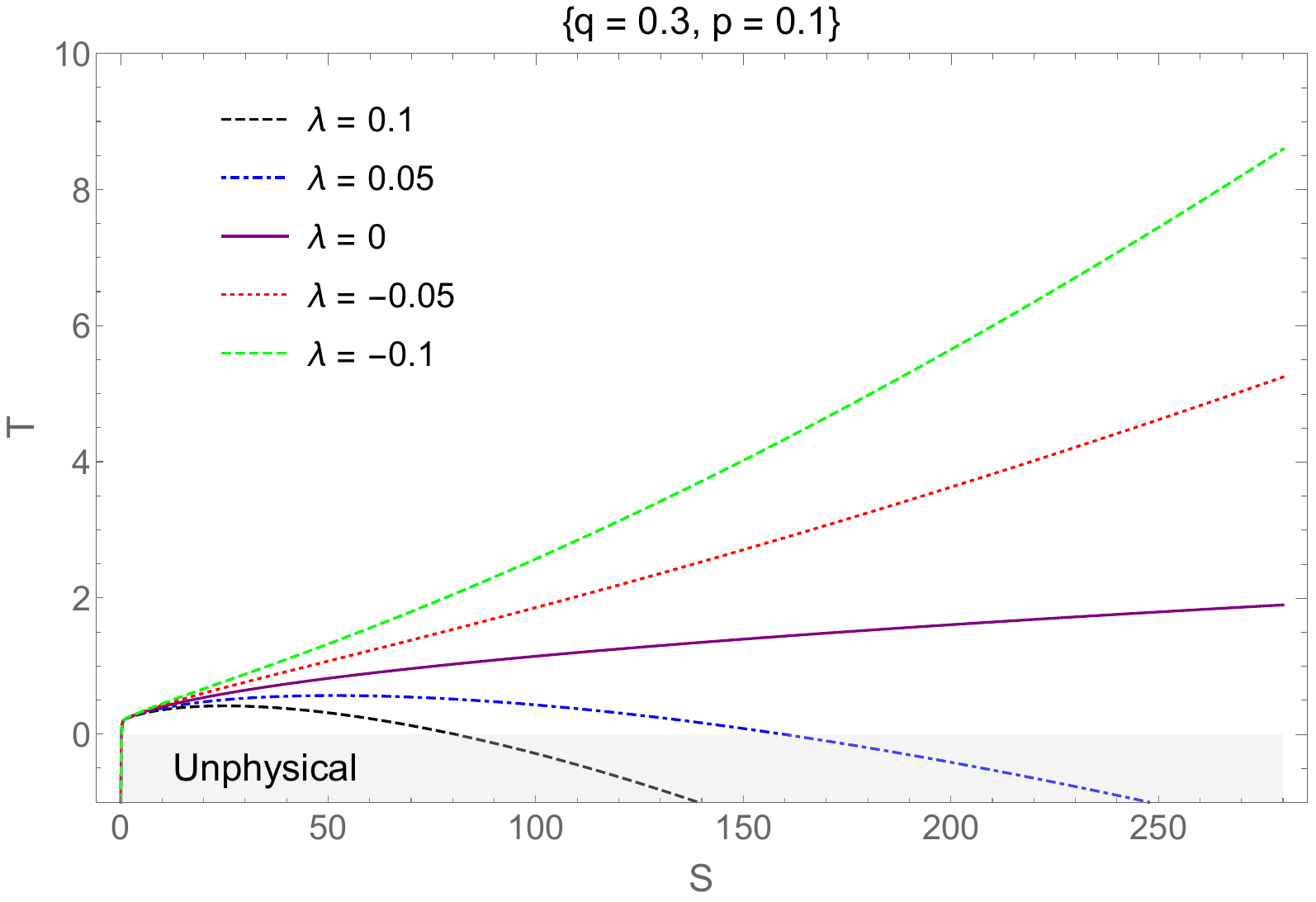}
\vspace{1mm}
\includegraphics[width=8cm]{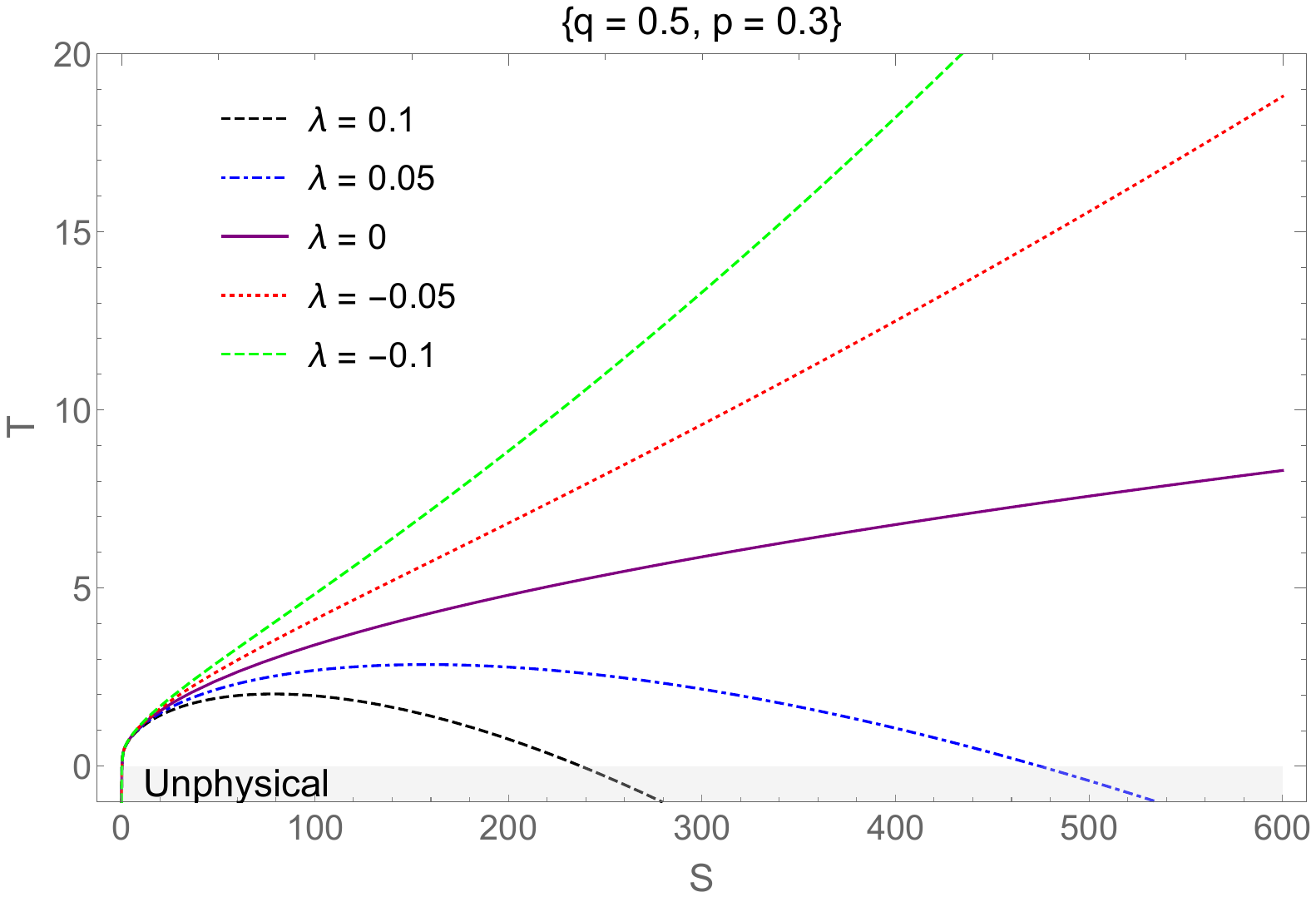}
\caption{Temperature of the charged AdS BH as a function of entropy, for various values of $\lambda$, and $(q,p)=(0.3, 0.1)$ (upper panel) and $(q,p)=(0.5, 0.3)$ (lower panel). The shaded region corresponds to the unphysical regime where $T<0$.}
\label{figTRN1}
\end{center}
\end{figure}

Now, by taking the partial derivative of the mass \( M \) with respect to the entropy \( S \), the temperature is obtained from Eq.~(\ref{MR2eq}) as
\begin{equation}
\label{TR1eq}
T_q(S) = \frac{\pi^2 S \left(1 + 8 p S\right) - \lambda S^3 - \pi^3 q^2}{4 \pi^{\frac{5}{2}} S^{\frac{3}{2}}}\,.
\end{equation}
The corresponding \( T\text{-}S \) diagram is shown in Fig.~\ref{figTRN1} for various values of \( \lambda \), \( p \) and \( q \). For \( \lambda \le 0 \), a non-physical regime (i.e., \( T < 0 \)) emerges at small values of \( S \), corresponding to the domain where \( M \) is a decreasing function of entropy. Excluding this region, the temperature is found to be a strictly increasing function of \( S \), indicating local thermodynamic stability for large charged AdS BHs.

In contrast, for \( \lambda > 0 \), two non-physical regions appear at both small and large values of \( S \), respectively. In the intermediate range, the temperature initially increases with \( S \), reaches a maximum, and then decreases as \( S \) continues to grow. As we shall see in the following analysis, this behavior corresponds to a transition from an initial phase of thermodynamic stability to a final regime in which the system becomes thermodynamically unstable.

\section{Thermodynamic Stability}
\label{Stab}

Since BHs exhibit thermodynamic properties, evaluating their stability is essential. In this section, we investigate the local and global thermodynamic stability of Schwarzschild-AdS and charged AdS BHs within the canonical ensemble. Local stability is examined through the behavior of the heat capacity, while global stability is assessed via the Gibbs free energy. These analyses provide valuable insights into the thermal behavior and phase structure of the corresponding BH solutions.



\subsection{Local stability}
The local stability of the BH is assessed by examining the sign of the specific heat at constant pressure, \( C_p \), which is defined as
\begin{equation}
\label{Cp}
C_p(S) = T\left(\frac{\partial S}{\partial T}\right)_p\,.
\end{equation}
A positive specific heat capacity (\( C_p > 0 \)) indicates local thermodynamic stability, implying that the system is stable under small thermal fluctuations. In contrast, a negative heat capacity (\( C_p < 0 \)) signals local instability, as the BH tends to amplify rather than damp thermal perturbations. Phase transitions are associated with divergences in \( C_p \), marking critical points where the system undergoes a transition between stable and unstable thermodynamic phases.

\subsubsection{Schwarzschild-AdS black hole}

\begin{figure}[t]
\begin{center}
\includegraphics[width=8cm]{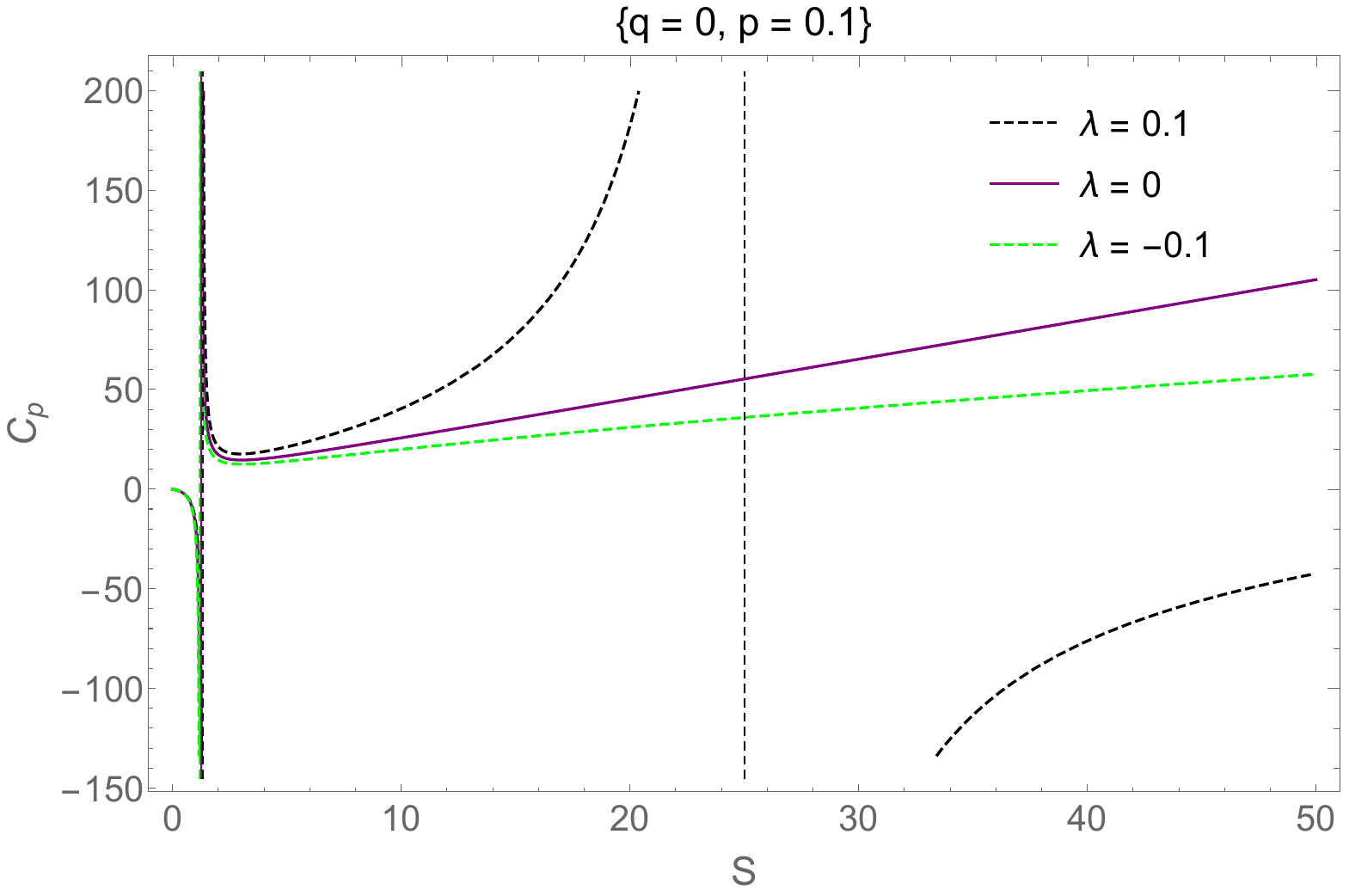}
\caption{Heat capacity of the Schwarzschild-AdS BH as a function of entropy, for various values of $\lambda$ and $p=0.1$.}
\label{figCS1}
\end{center}
\end{figure}

By taking partial derivative of Hawking temperature in Eq.~(\ref{TS1eq}) with respect to $ S $ at constant pressure $p$, we get
\begin{eqnarray}\label{CS}
C_p(S)=\frac{2S\left[\lambda S^2-\pi^2\left(1+8pS\right)\right]}{3\lambda S^2+\pi^2\left(1-8pS\right)}\,.
\end{eqnarray}
Figure~\ref{figCS1} illustrates the behavior of the heat capacity for Schwarzschild-AdS BHs as a function of entropy, evaluated for different values of the parameter \( \lambda \) at fixed pressure \( p = 0.1 \). For clarity of presentation, we restrict our analysis to three representative values \( \lambda = 0.1 \), \( 0 \), and \( -0.1 \).

Consistent with the analysis of the \( T\text{-}S \) profile, we observe that for \( \lambda \le 0 \), there is a single discontinuity corresponding to the transition point where the temperature curve exhibits zero slope (see Fig.~\ref{figTS}). Notably, this point lies within the physical domain where \( T > 0 \). Near this point, the heat capacity transitions from negative values at small entropy to positive values at larger entropy, thereby signaling a phase transition from a thermodynamic unstable small BH to a stable large one. 
In this context, the effect of the modified gravity correction is reflected in a slower growth of the heat capacity compared to the \( \lambda = 0 \) case.

Conversely, for \( \lambda > 0 \), two singularities appear. The first occurs at low entropy and corresponds to the same physical behavior previously described for \( \lambda \le 0 \). The second, however, is specific to this case and arises at large entropy. Near this singularity, the heat capacity transitions from positive to negative values as entropy increases. 
Accordingly, three distinct BH configurations can be identified: a small unstable BH with \( C_p < 0 \), an intermediate stable BH with \( C_p > 0 \), and a large unstable BH exhibiting once again \( C_p < 0 \). This structure represents a novel feature compared to the standard GR case and originates from the CKG  correction, which becomes relevant and affects the BH thermodynamic stability at very large scales.


\subsubsection{Charged AdS black hole}

\begin{figure}[t]
\begin{center}
\includegraphics[width=8cm]{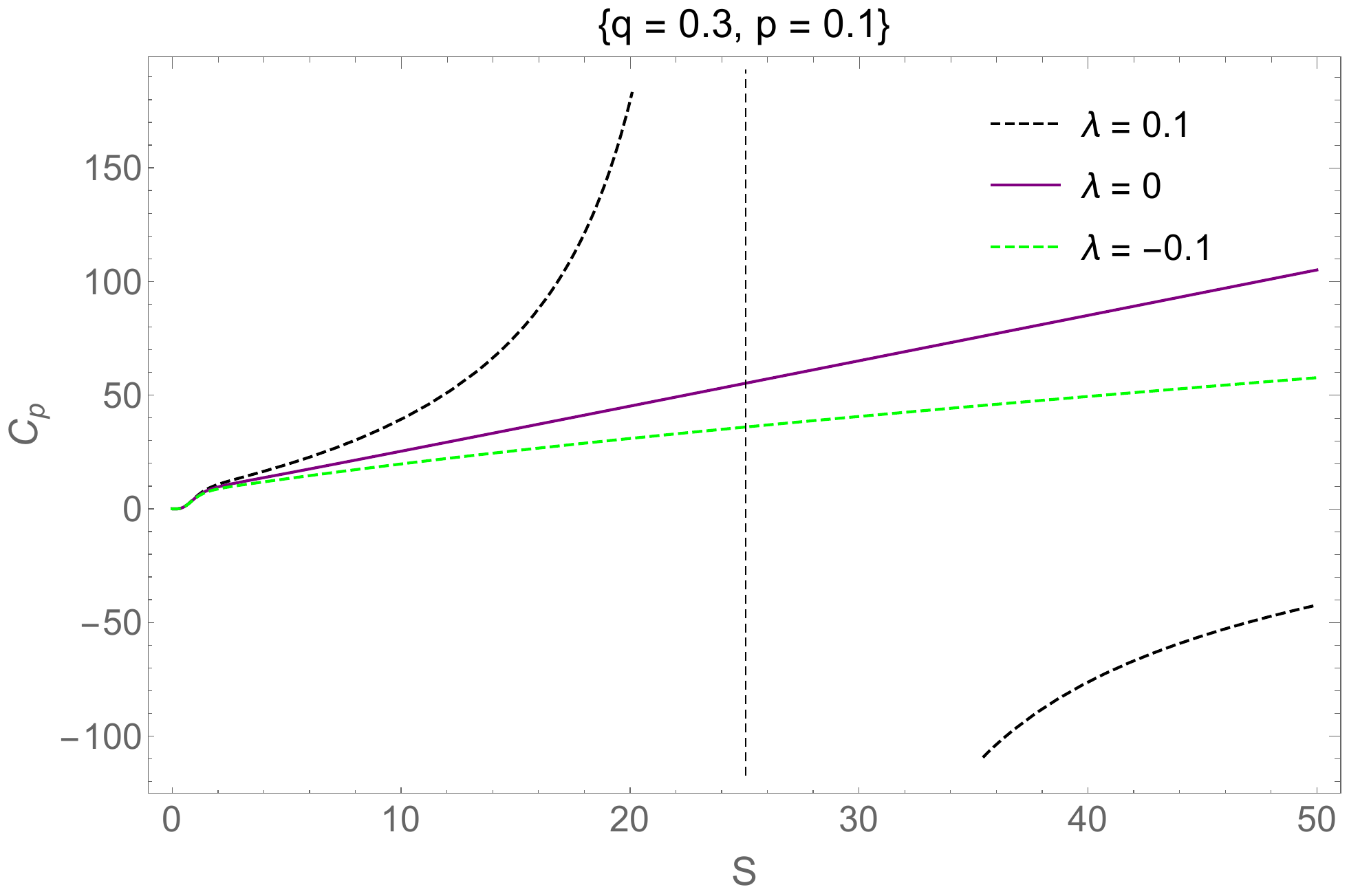}
\vspace{1mm}
\includegraphics[width=8cm]{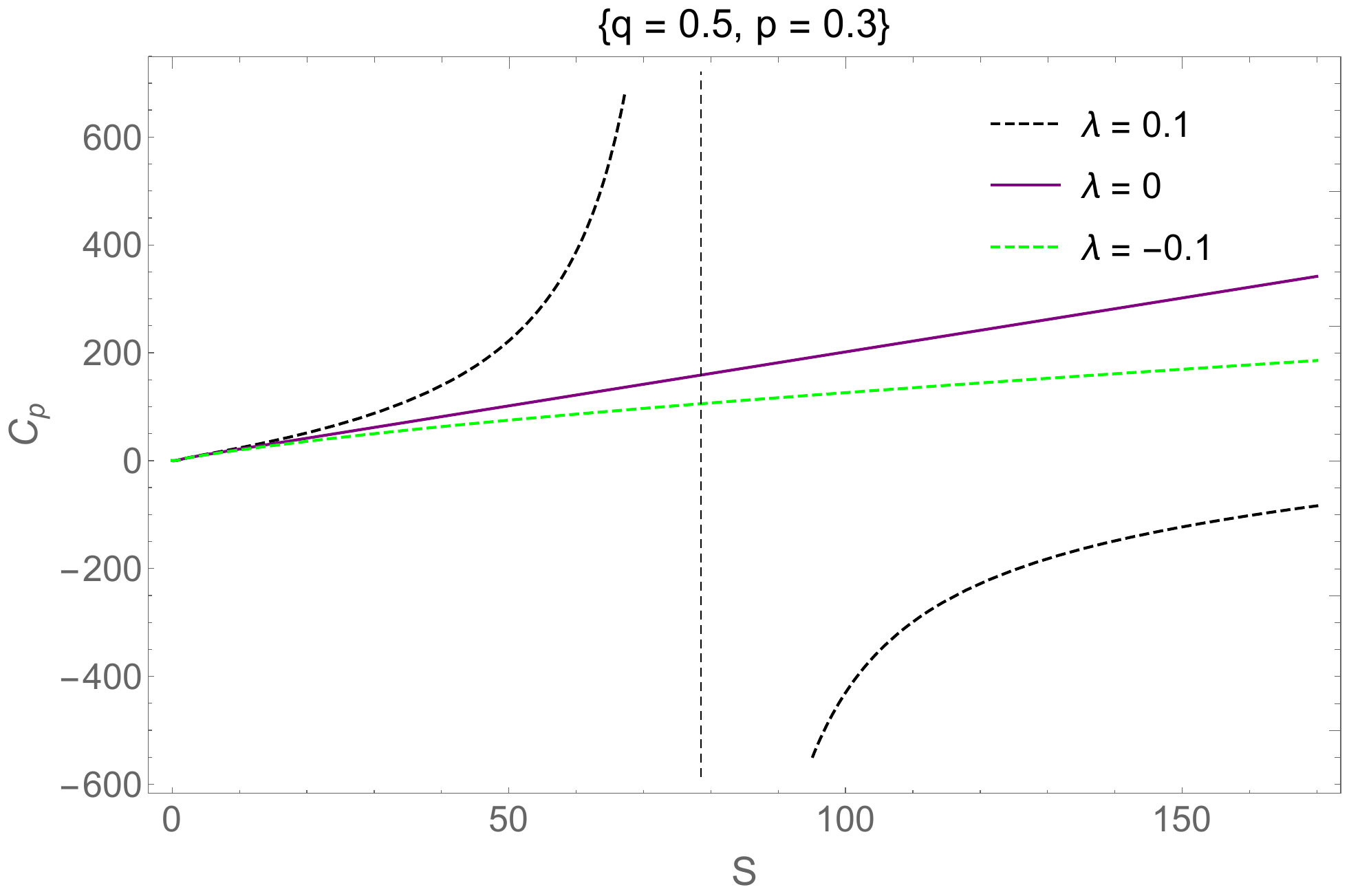}
\caption{Heat capacity of the charged AdS BH as a function of entropy, for various values of $\lambda$, and $(q,p)=(0.3, 0.1)$ (upper panel) and $(q,p)=(0.5, 0.3)$ (lower panel).}
\label{figCRN1}	
\end{center}
\end{figure}

From Eqs.~(\ref{TR1eq}) and~(\ref{Cp}), the heat capacity of the charged AdS BH can be expressed as follows:
\begin{equation}
\label{CRN}
C_{q,p}(S)=\frac{2S\left\{\lambda S^3+\pi^2 \left[\pi q^2-S\left(1+8pS\right)\right]\right\}}
{3\lambda S^3+\pi^2\left[S\left(1-8pS\right)-3\pi q^2\right]}
\,.
\end{equation}
Figure~\ref{figCRN1} illustrates the behavior of $C_p$ as a function of entropy $S$ for charged AdS black holes under various values of the parameters $\lambda$, $p$, and $q$. Unlike the previous case, we observe that the presence of a non-zero charge causes the heat capacity to be either a continuous function of entropy (for $\lambda \leq 0$), or to exhibit at most a unique singularity (for $\lambda > 0$). In the former case, excluding the unphysical region where $T < 0$ (corresponding to very small BH sizes, see Fig.~\ref{figTRN1}), the positive sign of $C_p$ indicates a thermodynamically stable configuration of the charged BH. Conversely, for $\lambda > 0$, the heat capacity undergoes a transition from positive values at small $S$ to negative values at large $S$, signaling a transition from a stable small BH configuration to an unstable large BH configuration.


\subsection{Gibbs free energy and global stability}
In the canonical ensemble, the global stability of a thermodynamic system, such as a BH, is evaluated through the Gibbs free energy, \( G \). At thermal equilibrium, the system naturally evolves toward minimizing \( G \). A negative Gibbs free energy (\( G < 0 \)) indicates that the system is in a thermodynamically favorable state relative to a specified reference background, typically pure AdS spacetime in gravitational contexts. Accordingly, within the canonical ensemble, a BH configuration with lower (more negative) \( G \) is considered globally stable and less susceptible to undergoing phase transitions or large-scale thermodynamic fluctuations.

\subsubsection{Schwarzschild-AdS black hole}

\begin{figure}[t]
\begin{center}
\includegraphics[width=8cm]{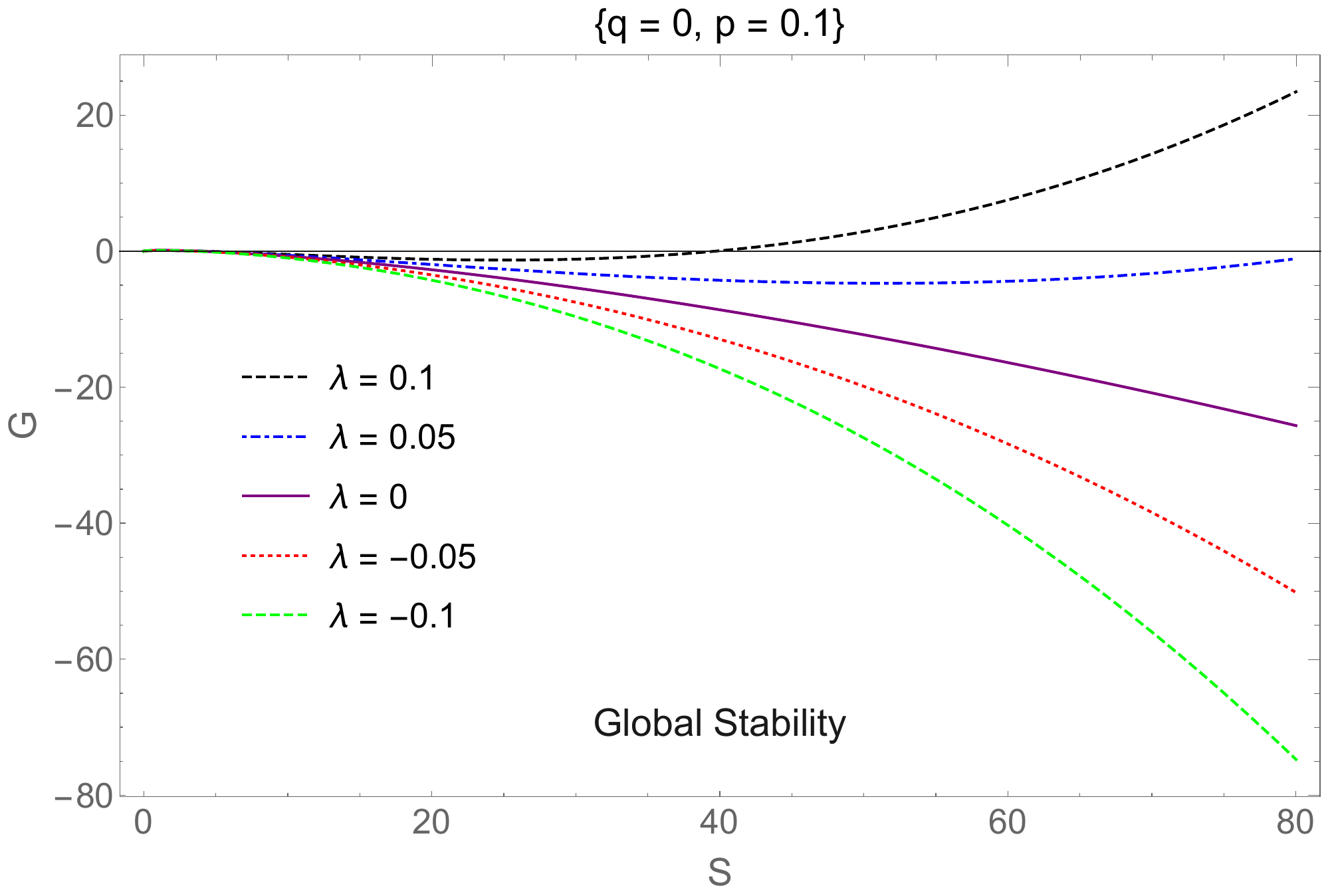}
\caption{Free energy of the Schwarzschild-AdS BH as a function of entropy, for various values of  $\lambda$ and $p=0.1$.}
\label{figGS1}
	\end{center}
\end{figure}
The Gibbs free energy is defined as
\begin{equation}
G=M-TS\,.
\end{equation}
By employing the expressions for the mass and temperature derived for Schwarzschild-AdS, we straightforwardly obtain
\begin{equation}
G(S)=\frac{\sqrt{S}\left[5\pi^2\left(3-8pS\right)+9\lambda S^2\right]}{60\pi^{5/2}}\,.
\end{equation}
The Gibbs free energy as a function of entropy is shown for Schwarzschild-AdS Bhs in Fig.~\ref{figGS1}. Once again, we observe markedly different behaviors depending on the sign of the CKG parameter. Focusing on the physically relevant regime where the temperature remains positive (\( T > 0 \)), it is evident that negative values of \( \lambda \) correspond to a lower Gibbs free energy compared to the GR case with \( \lambda = 0 \). Notably, the more negative the value of \( \lambda \), the more pronounced this behavior becomes. 

Physically, this suggests that for \( \lambda < 0 \), the thermodynamic system is energetically more favorable, potentially reflecting enhanced stability or the presence of additional microscopic degrees of freedom that lower the free energy. Such a decrease in Gibbs free energy may also indicate a modification in the effective interactions or a suppression of long-range correlations, consistent with a richer thermodynamic landscape induced by the underlying modified gravity theory (see the discussion in Sec.~\ref{Geo}).

Conversely, for \( \lambda > 0 \), the Gibbs free energy is found to be greater than in GR. Interestingly, in this scenario, \( G \) is negative for relatively small entropy values, transitioning to positive values as \( S \) increases. The rate of this increase becomes more pronounced as \( \lambda \) grows. This sign change in the Gibbs free energy is reminiscent of a Hawking-Page-like transition, where a BH phase becomes less thermodynamically favorable than thermal AdS space as the system evolves. In this context, the modified gravity contribution appears to enhance this effect, potentially amplifying the role of geometric corrections in determining phase stability.

\subsubsection{Charged AdS black hole}

\begin{figure}[t]
\begin{center}
\includegraphics[width=8cm]{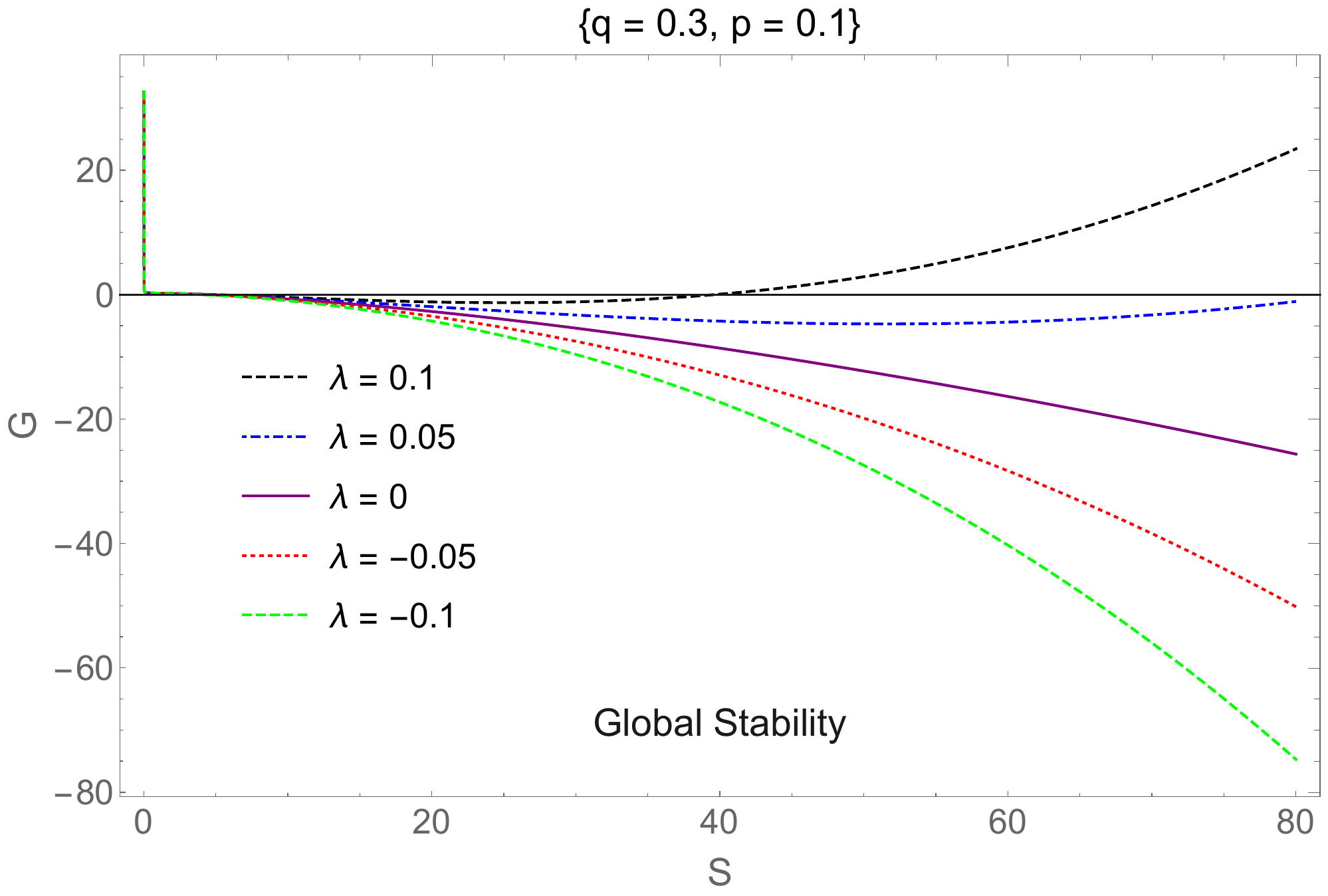}
\vspace{1mm}
\includegraphics[width=8cm]{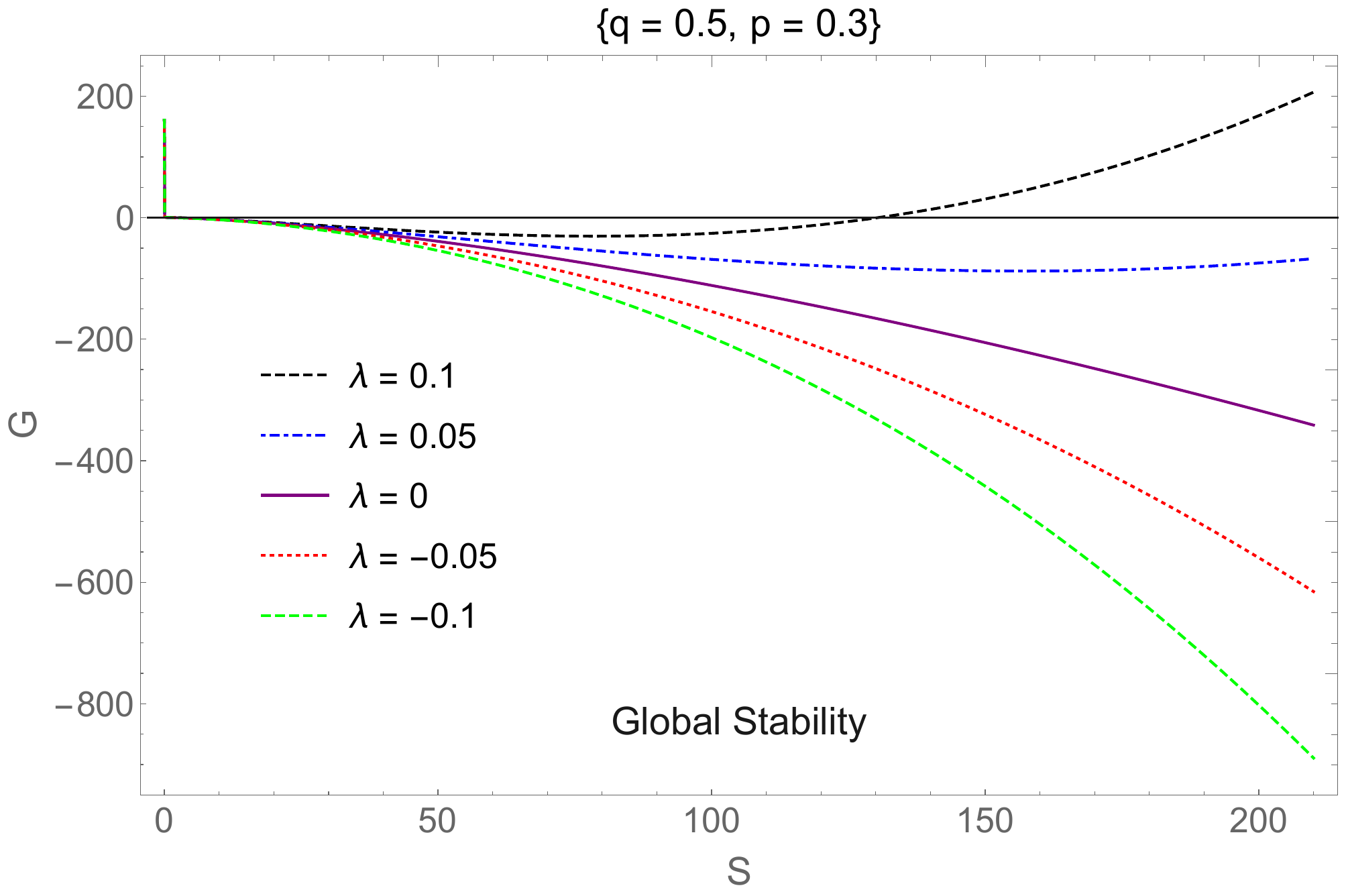}
\caption{Free energy of the charged AdS BH as a function of entropy, for various values of $\lambda$, and $(q,p)=(0.3, 0.1)$ (upper panel) and $(q,p)=(0.5, 0.3)$ (lower panel).}
\label{figGRN1}
	\end{center}
\end{figure}

The Gibbs free energy for charged AdS BHs takes the generalized form
\begin{equation}
    G_q(S) = \frac{5\pi^2\left[S \left(3 - 8pS\right) + 9\pi q^2\right] + 9\lambda S^3}{60\pi^{5/2}\sqrt{S}}\,,
\end{equation}
whose behavior as a function of entropy \( S \) is illustrated in Fig.~\ref{figGRN1} for different values of the parameters \( \lambda \), \( p \), and \( q \). The profile resembles that observed in the Schwarzschild-AdS case. The key difference, however, lies in the presence of a nonzero charge, which ensures that the Gibbs free energy remains positive at small entropy values, regardless of the sign of \( \lambda \). 

Physically, this behavior can be attributed to the additional contribution to the energy from the electromagnetic field, which becomes dominant in the small-\( S \). In this limit, the electrostatic energy stored in the BH scales inversely with its size and thus contributes significantly to the total free energy. As a result, the thermodynamic cost of sustaining a charged BH at small horizon radii is higher, leading to a positive Gibbs free energy even in cases where the uncharged counterpart would be energetically favored. 


\section{Geometrothermodynamic properties}
\label{Geo}

Since BHs can be assigned a well-defined temperature, it is natural to consider the existence of an associated microscopic structure. In recent years, considerable attention has been devoted to exploring the possible microstructure and underlying interactions of BHs~\cite{Cai:1998ep,Wei:2015iwa,Wei:2019uqg,Guo:2019oad,Xu:2020gud,Ghosh:2020kba,Xu:2020ftx,Prom,Dehghani:2023yph,Luciano:2023fyr,Luciano:2023bai,Ghaffari:2023vcw,Sekhmani:2024udl}. These studies suggest that BHs may possess a microscopic composition analogous to that of molecules in a non-ideal fluid.

To probe the nature of interactions among these hypothetical BH microconstituents, a widely adopted approach is the application of thermodynamic geometry. In this framework, the entire macroscopic thermodynamic system is endowed with a geometric structure that captures information about its internal statistical behavior. In particular, the geometric formalisms developed by Weinhold~\cite{Wein1} and Ruppeiner~\cite{Rupp1,Rupp2} have proven effective in revealing qualitative features of microscopic interactions in standard thermodynamic systems.

The curvature of the corresponding thermodynamic metric offers insight into the dominant type of interaction. A negative scalar curvature typically signals attractive interactions among microstructures, while a positive curvature indicates repulsion. A vanishing curvature, on the other hand, corresponds to either a lack of interactions - as in an ideal gas - or to a perfect balance between attractive and repulsive forces.

To assess the extent to which CKG impacts BH geometrothermodynamics, 
we now compute the Weinhold and Ruppeiner scalar curvatures based on the outcomes of Sec.~\ref{Ther}. In this setting, the Weinhold metric is defined as the Hessian (second derivative) of the internal energy with respect to the chosen thermodynamic variables~\cite{Wein1}. For BHs, identifying the internal energy with the mass~\eqref{firstlaw}, the metric components are given by:
\begin{eqnarray}
\label{gw}
g_{ij}^w = -\partial_i \partial_j M(S, p, q) \quad \Longrightarrow \quad ds^2_w = g^w_{ij} dx^i dx^j\,,
\end{eqnarray}
where \( x^i \) denotes a set of independent thermodynamic fluctuation coordinates.

Similarly, in the Ruppeiner formalism, the entropy is treated as the fundamental thermodynamic potential. Accordingly, the Ruppeiner metric is defined as the negative Hessian of the entropy with respect to the extensive thermodynamic variables:
\begin{equation}
\label{R1}
g_{ij}^{\text{Rup}} = -\partial_i \partial_j S\,.
\end{equation}
From Eqs.~\eqref{gw} and \eqref{R1}, together with the definition of temperature, it follows that the Weinhold and Ruppeiner line elements are related by a conformal transformation, where the temperature serves as the conformal factor:
\begin{equation}
\label{R2}
ds^2_R = \frac{ds^2_w}{T}\,.
\end{equation}

\subsection{Schwarzschild-AdS black hole}

We now turn our attention to the geometrothermodynamic analysis of Schwarzschild-AdS BHs based on Eq.~\eqref{R2}. The preference for the Ruppeiner geometry over the Weinhold framework is justified by its well-established connection to fluctuation theory in statistical mechanics~\cite{Rupp1,Rupp2}, which endows it with a more direct physical interpretation. Furthermore,  we work in the canonical (fixed charge) ensemble, considering the entropy \( S \) and pressure $p$ as the relevant thermodynamic fluctuation coordinates. 

\begin{figure}[t]
\begin{center}
\includegraphics[width=8cm]{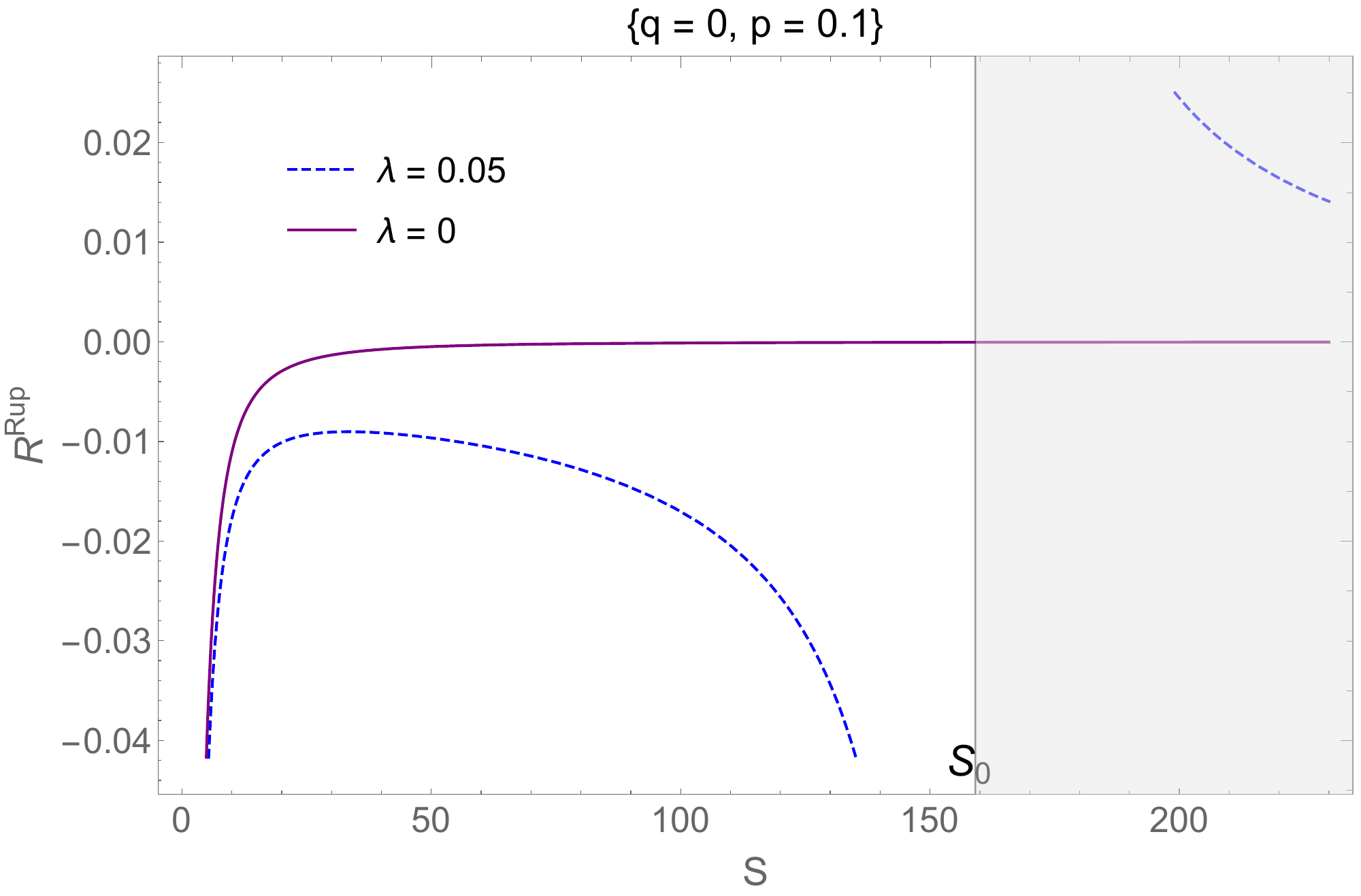} 

\vspace{0.5cm}
\includegraphics[width=8cm]{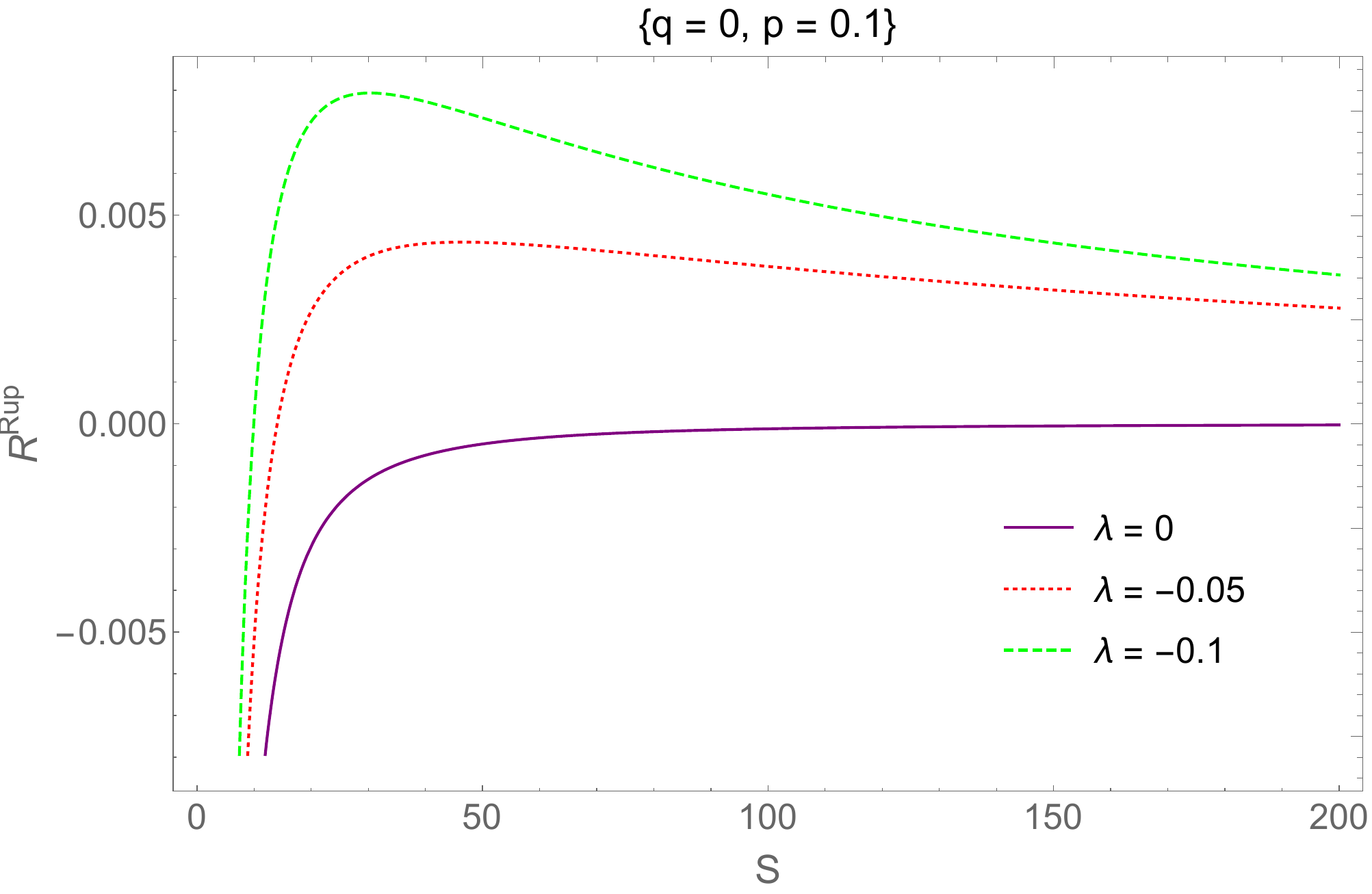} 
\caption{Ruppeiner curvature of the Schwarzschild-AdS BH as a function of entropy, for $\lambda > 0$ (upper panel) and $\lambda<0$ (lower panel). The shaded region corresponds to the unphysical regime where $T<0$. 
}
\label{RRup1}
\end{center}
\end{figure}

Under these assumptions, 
the Ruppeiner scalar curvature takes the following form:
\begin{equation}
\label{Rscalar}
\mathcal{R}^{Rup} (S,p) = \frac{\lambda S^2+\pi^2}{\lambda S^3-\pi^2 S\left(1+8pS\right)}\,.
\end{equation}
The behavior of \( R^{\mathrm{Rup}} \) as a function of entropy is illustrated in Fig.~\ref{RRup1} for $\lambda>0$ (upper panel) and $\lambda<0$ (lower panel). For the selected set of model parameters, it is evident that the correction introduced by the \( \lambda \)-term significantly modifies the behavior of the Ruppeiner curvature when compared to the reference configuration \( \lambda = 0 \) (purple curve). In this latter case, \( R^{\mathrm{Rup}} \) remains strictly negative, evolving from large negative values for small BH sizes to values asymptotically approaching zero at large entropy \( S \). According to the standard interpretation of Ruppeiner curvature, this behavior suggests that at large entropy, the BH approximates an ideal gas, characterized by negligible interactions among its microscopic constituents.  As the black hole evaporates and shrinks, these interactions intensify and become predominantly attractive (\( R_{\mathrm{Rup}} < 0 \)). This trend indicates a transition in the microscopic structure of the system, wherein  correlations among the underlying degrees of freedom grow increasingly significant as the Planckian regime is approached.



A qualitatively similar behavior is observed for  \( \lambda > 0 \) (blue dashed curve, upper panel of Fig.~\ref{RRup1}). However, in this case, the analytical expression \eqref{Rscalar} reveals that \( R^{\mathrm{Rup}} \) exhibits an apparent singularity at a specific entropy value \( S_0 \). This singularity, however, coincides with the physical limiting point, i.e., the point at which the temperature vanishes, as can be directly inferred from Eqs.~\eqref{TS1eq} and~\eqref{R2}. Consequently, only the region \( S < S_0 \), where the absolute temperature remains strictly positive, should be regarded as physically meaningful.

Furthermore, by comparing with the case \( \lambda = 0 \), we observe that although the Ruppeiner curvature remains negative throughout the physical domain, its magnitude is consistently greater, particularly for relatively large values of \( S \). This indicates the presence of stronger attractive interactions among the microscopic constituents of the BH. This result suggests that the modified dynamics in CKG enhances the correlations among the BH's microscopic degrees of freedom, leading to stronger effective attractive interactions. 

On the other hand, for \( \lambda < 0 \) (lower panel of Fig.~\ref{RRup1}), it can be seen that the scalar curvature does not exhibit any singularities for positive values of entropy \( S \). Moreover, compared to the prediction from GR, the behavior exhibits a much richer phenomenology: while \( R^{\mathrm{Rup}} \) asymptotically approaches zero at high entropy and becomes strongly negative at low \( S \) (similarly to Einstein's theory), it notably features an intermediate region where the curvature is positive. This indicates the emergence of repulsive interactions among the microscopic constituents of the BH within that entropy range, leading to qualitatively distinct interaction patterns that are not observed in standard GR.

\begin{figure}[t]
\begin{center}
\includegraphics[width=8cm]{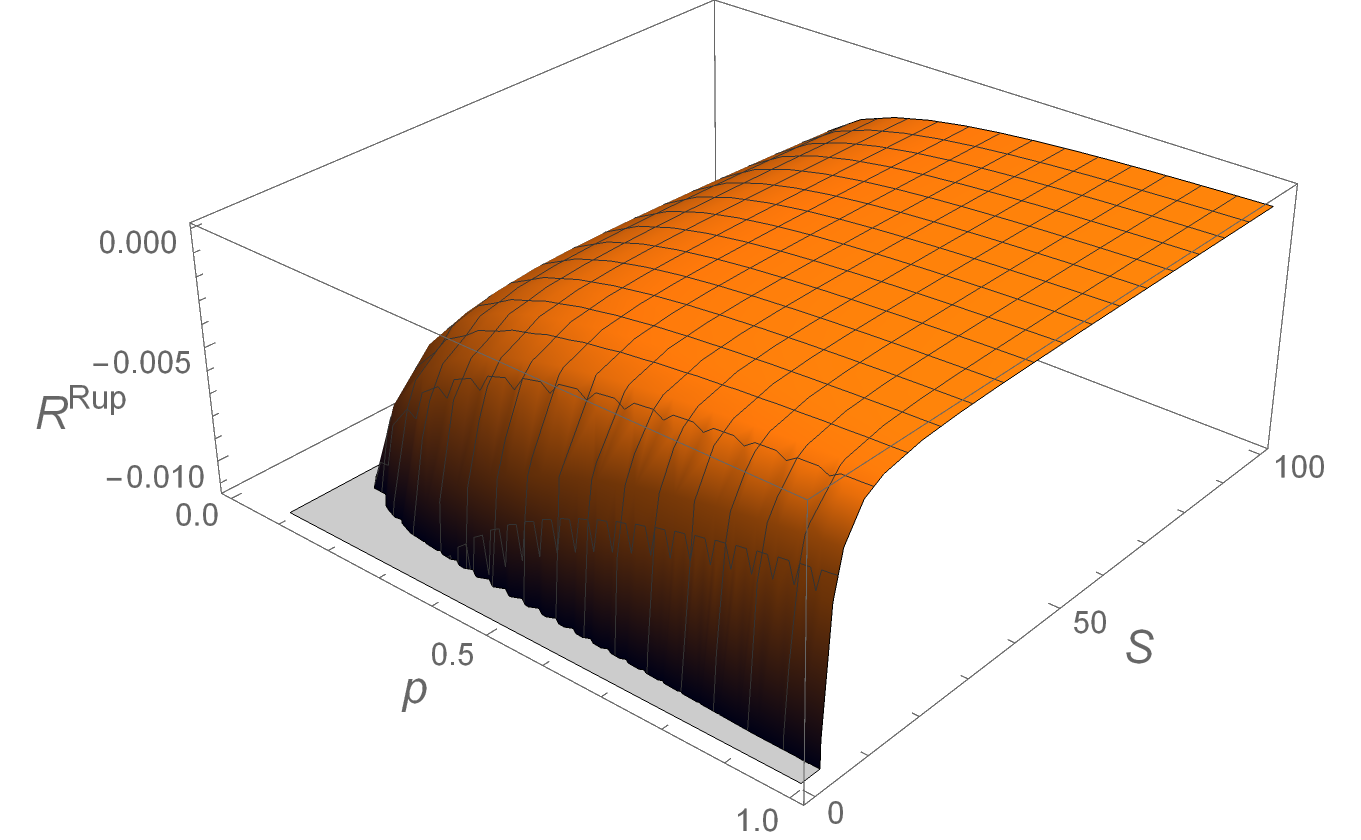} 

\vspace{0.5cm}
\includegraphics[width=8cm]{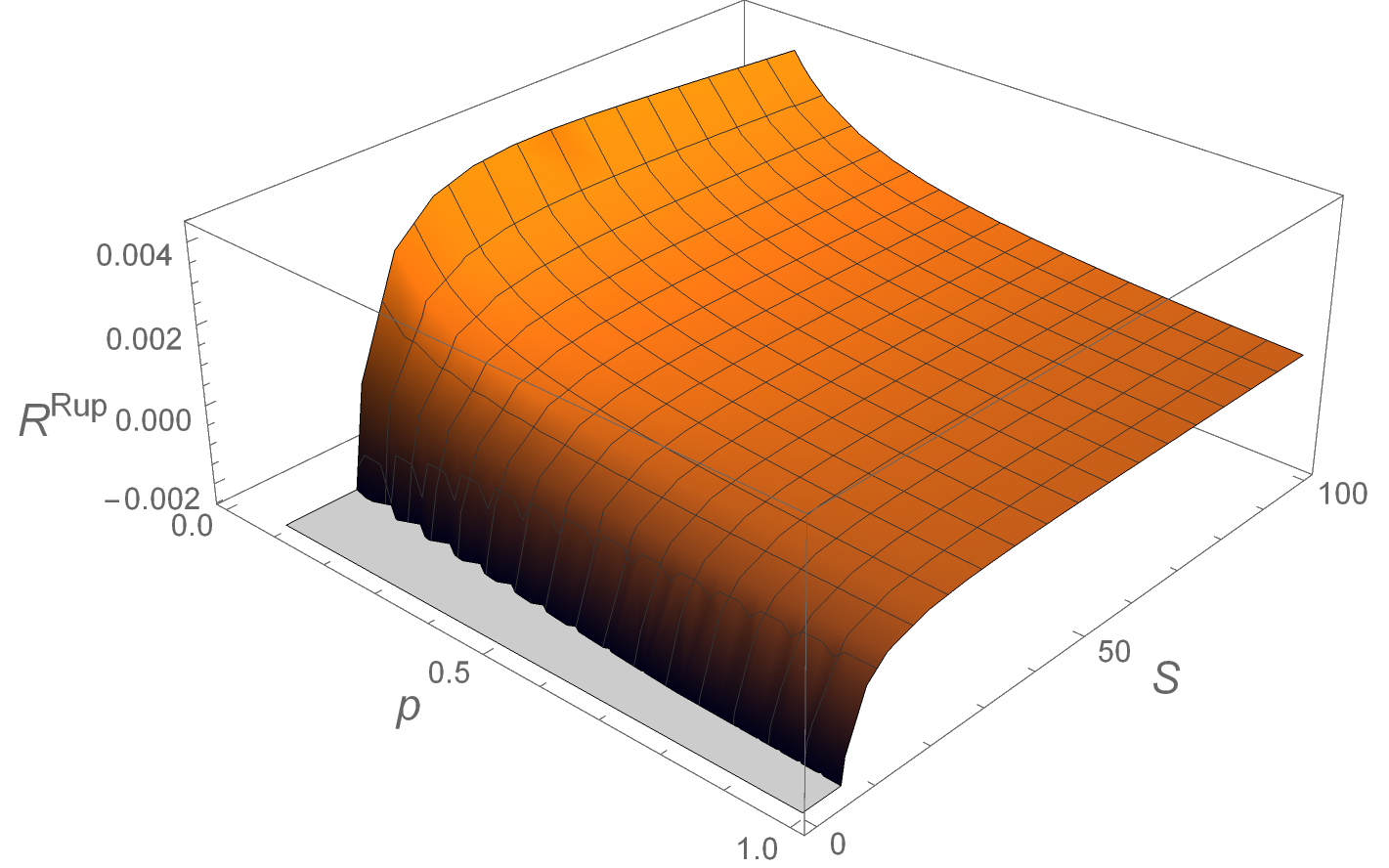} 
\caption{3D plot of Ruppeiner curvature of Schwarzschild-AdS BH as a function of entropy and pressure, for $\lambda = 0.05$ (upper panel) and $\lambda=-0.05$ (lower panel).
}
\label{3D1}
\end{center}
\end{figure}

The fact that only negative $\lambda$ gives rise to such a repulsive regime suggests that the CKG modification, when introduced with negative coupling, may soften the attractive nature of gravitational interactions at intermediate/large scales, possibly as a consequence of underlying competition between classical and non-classical (higher-derivative) contributions to the thermodynamic potential.
Physically, this may be interpreted as a signature of effective microstructural screening mechanisms that become relevant only when the modified gravity term counteracts (rather than reinforces) the standard Einstein-Hilbert dynamics. As a result, the system exhibits behavior reminiscent of repulsive statistical forces - similar to those seen in fermionic systems or other condensed matter analogues~\cite{Mann,Cai:2013qga,Hendi:2012um,Mo:2014qsa,Li:2014ixn,Xu:2014kwa,Mo:2013ela,Hansen:2016ayo,Vedral} - before transitioning back into an attractively interacting phase as the BH continues to evaporate and its size approaches zero.

The features discussed above, including the behavior of the curvature as a function of pressure, are further illustrated in the 3D plots shown in Fig.~\ref{3D1}, for \( \lambda > 0 \) (upper panel) and \( \lambda < 0 \) (lower panel).

\subsection{Charged AdS black hole}
Let us now consider the case of charged BHs. For such systems, the Ruppeiner scalar curvature takes the following form:
\begin{equation}
\label{curv2}
    \mathcal{R}^{Rup}_q (S,p) = \frac{\lambda S^3+\pi^2 S-2\pi^3 q^2}{\lambda S^4+\pi^2 S\left[\pi q^2-S\left(1+8pS\right)\right]}\,.
\end{equation}
It is straightforward to verify that for $\lambda = 0$, the standard expression for RN-AdS BHs is recovered~\cite{Guo:2019oad}.

\begin{figure}[t]
\begin{center}
\includegraphics[width=8cm]{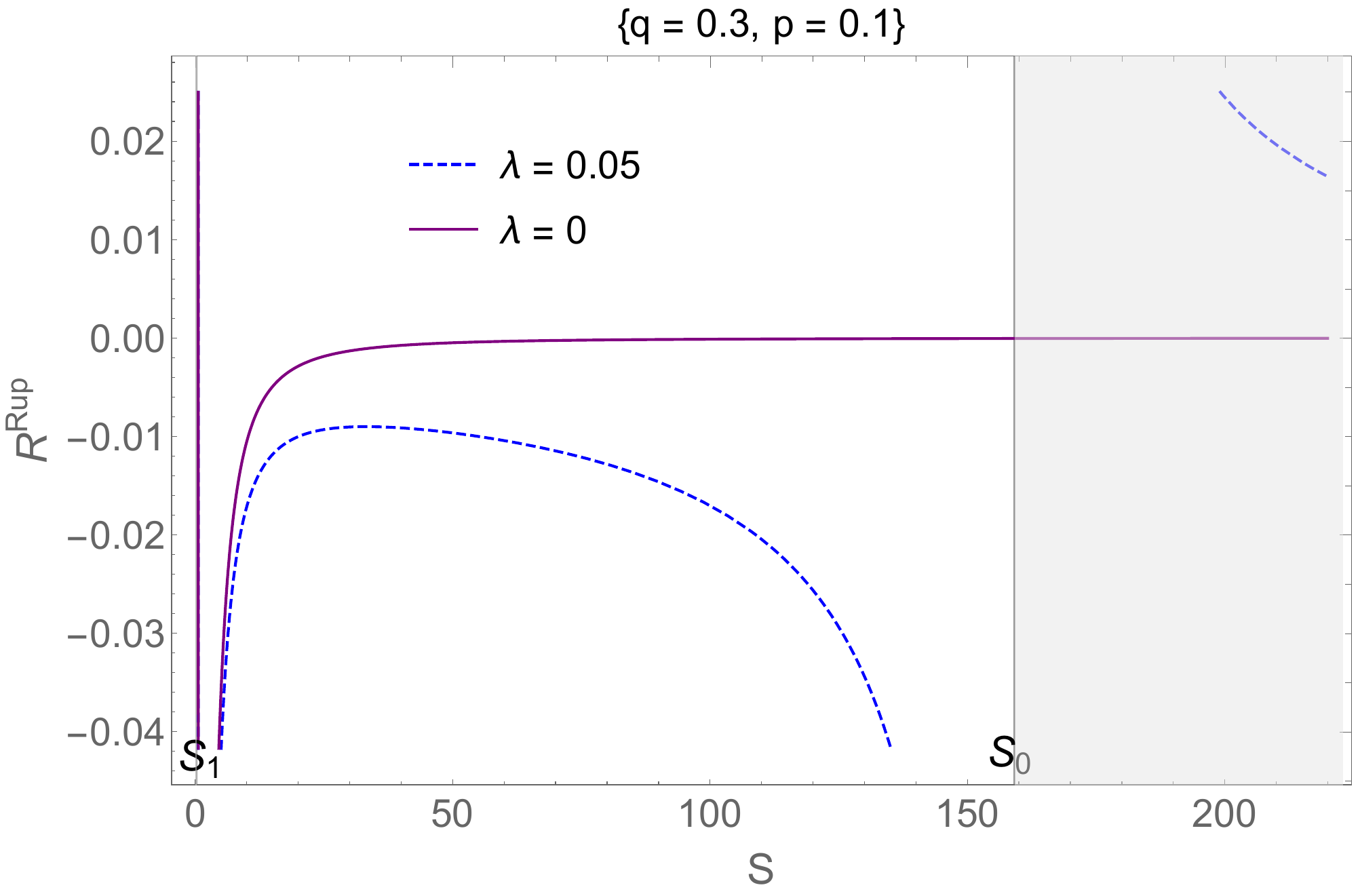} 

\vspace{0.5cm}
\includegraphics[width=8cm]{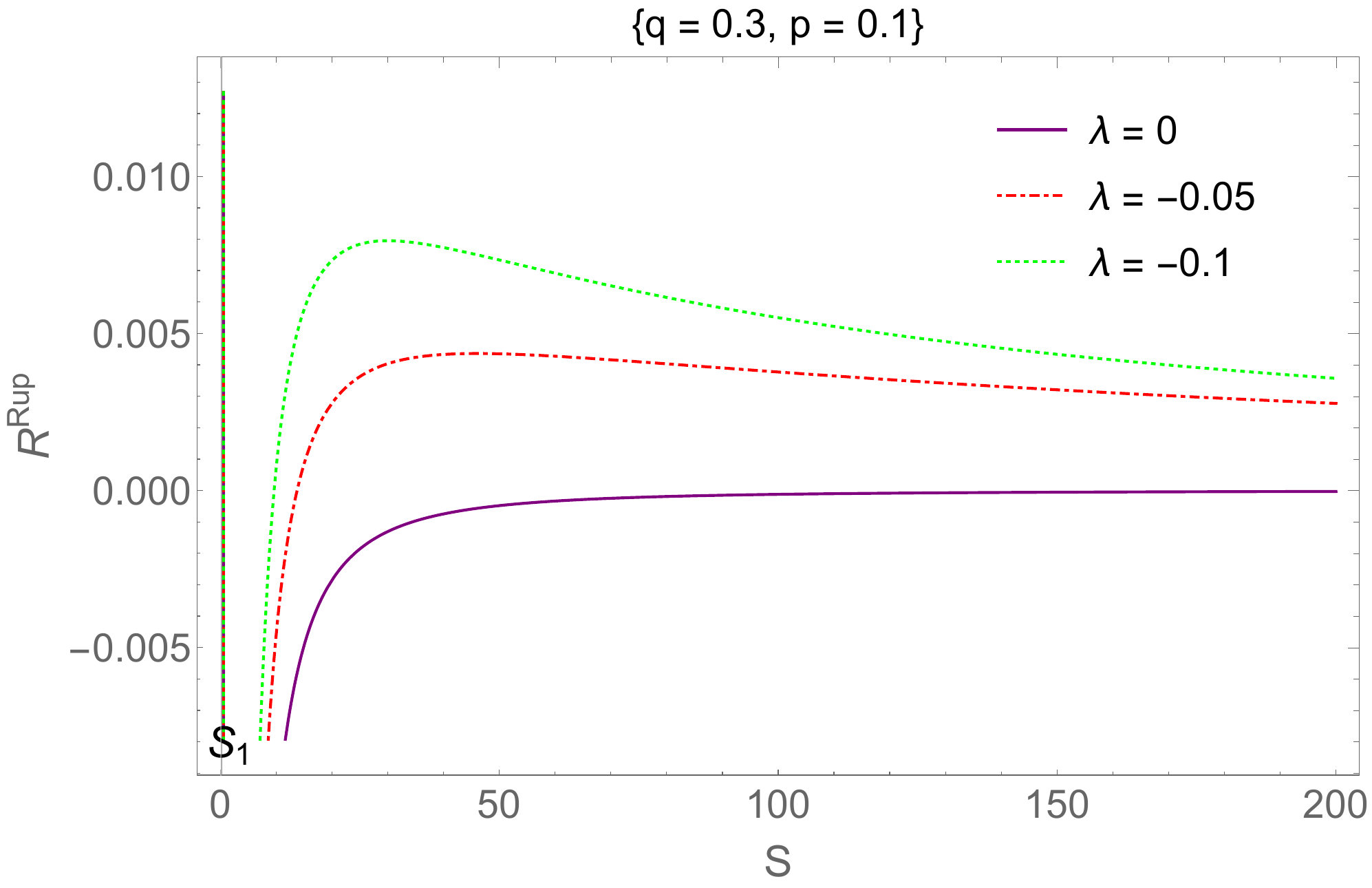} 
\caption{Ruppeiner curvature of the charged AdS BH as a function of entropy, for $\lambda > 0$ (upper panel) and $\lambda<0$ (lower panel). The shaded regions correspond to the unphysical regime where $T<0$. 
}
\label{RRup2}
\end{center}
\end{figure}

The behavior of the curvature~\eqref{curv2} is displayed in Fig.~\ref{RRup2} for \(\lambda > 0 \) (upper panel) and \( \lambda < 0 \) (lower panel). 
A comparison with the corresponding plots in Fig.~\ref{RRup1} shows that an additional (\( q \)-) divergence arises, denoted by \( S_1 \), which corresponds to another physical limitation point. As a result, the physically admissible range of entropy is bounded from below for all values of \( \lambda \), while an upper bound exists only in the case \( \lambda > 0 \), consistent with the Schwarzschild-AdS analysis.

Interestingly, as \( S \rightarrow S_1^+ \), the Ruppeiner curvature is positive. This feature is entirely attributable to the presence of charge, as it has no analogue in the Schwarzschild-AdS BH scenario. As the entropy increases, however, the influence of the charge contribution becomes subdominant, and the behavior of the curvature becomes qualitatively similar to that observed in the Schwarzschild-AdS case. Such features
highlight a nontrivial interplay between the effects of charge and the corrections induced by CKG,  suggesting that the microstructure of BHs are governed by a more intricate balance of geometric and charge degrees of freedom, especially in near-extremal configurations \cite{Harada,Harada2a}.

\begin{figure}[t]
\begin{center}
\includegraphics[width=8cm]{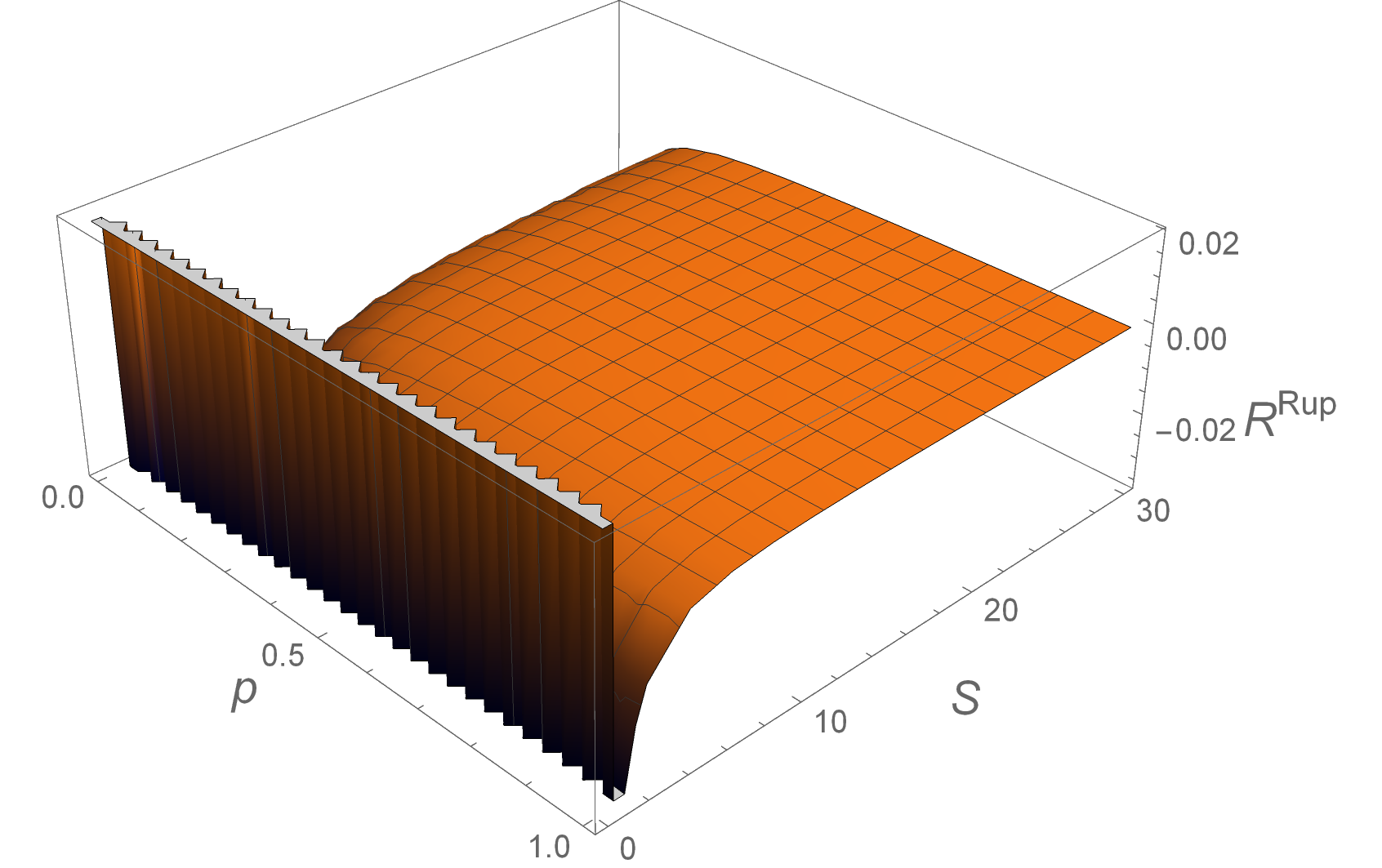} 

\vspace{0.5cm}
\includegraphics[width=8cm]{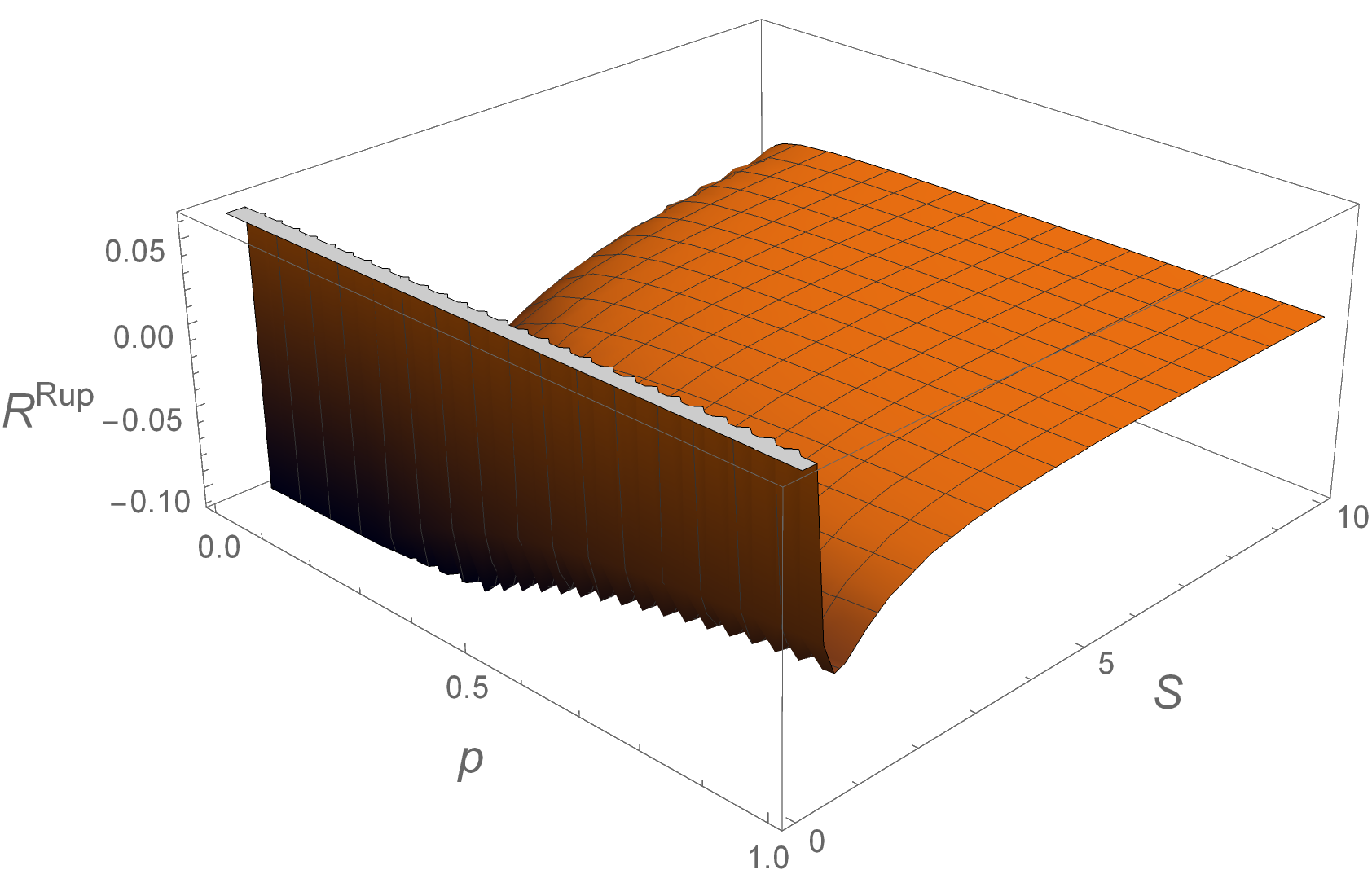} 
\caption{3D plot of Ruppeiner curvature of charged AdS BH as a function of entropy and pressure, for fixed $q=0.3$, $\lambda = 0.05$ (upper panel) and $\lambda=-0.05$ (lower panel).
}
\label{3D2}
\end{center}
\end{figure}

A more comprehensive perspective on the aforementioned characteristics can be gained by examining the 3D plots presented in Fig.~\ref{3D2}, corresponding to \( \lambda > 0 \) (upper panel) and \( \lambda < 0 \) (lower panel).

\begin{figure}[t]
\begin{center}
\includegraphics[width=8cm]{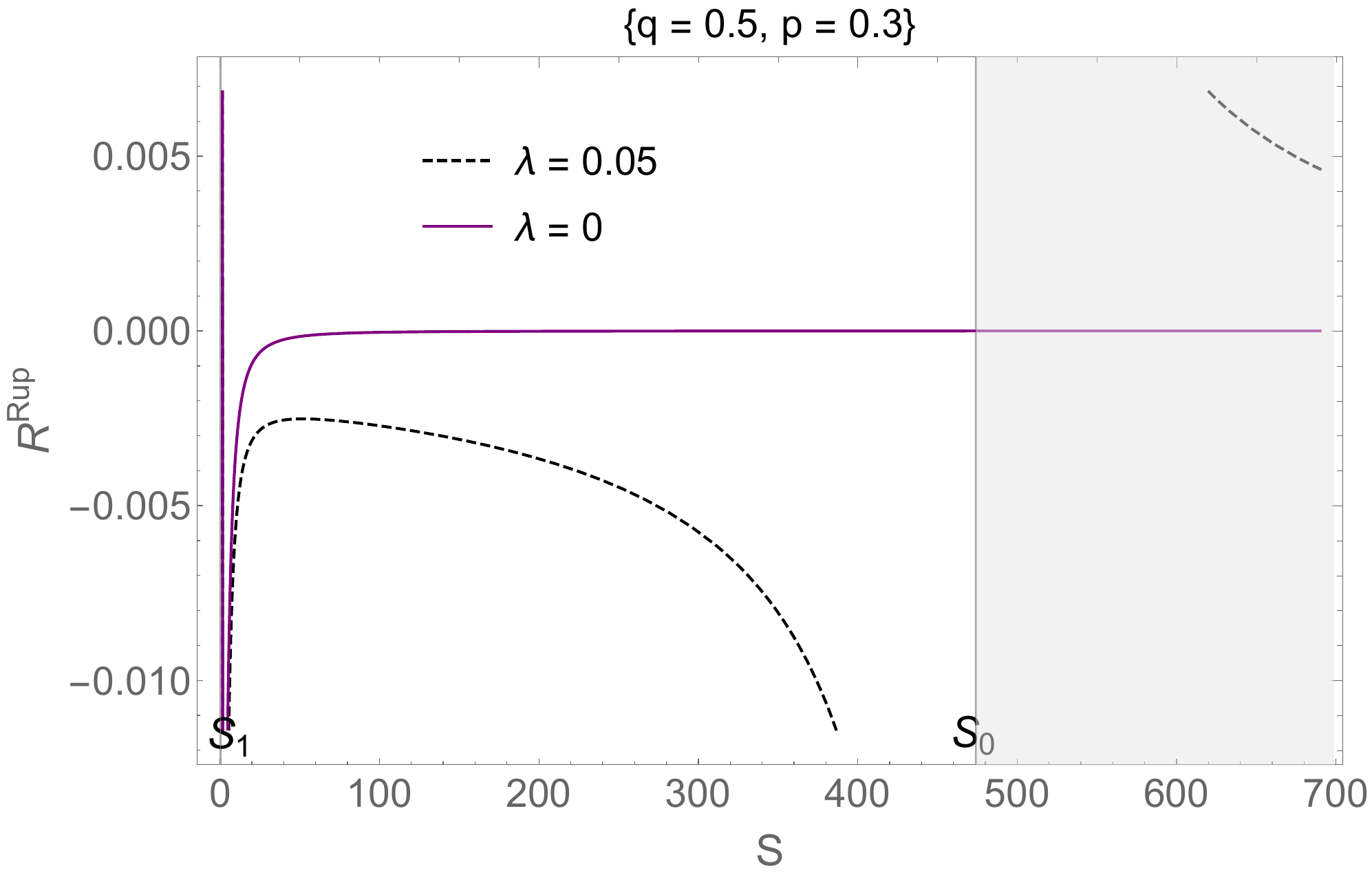} 

\vspace{0.5cm}
\includegraphics[width=8cm]{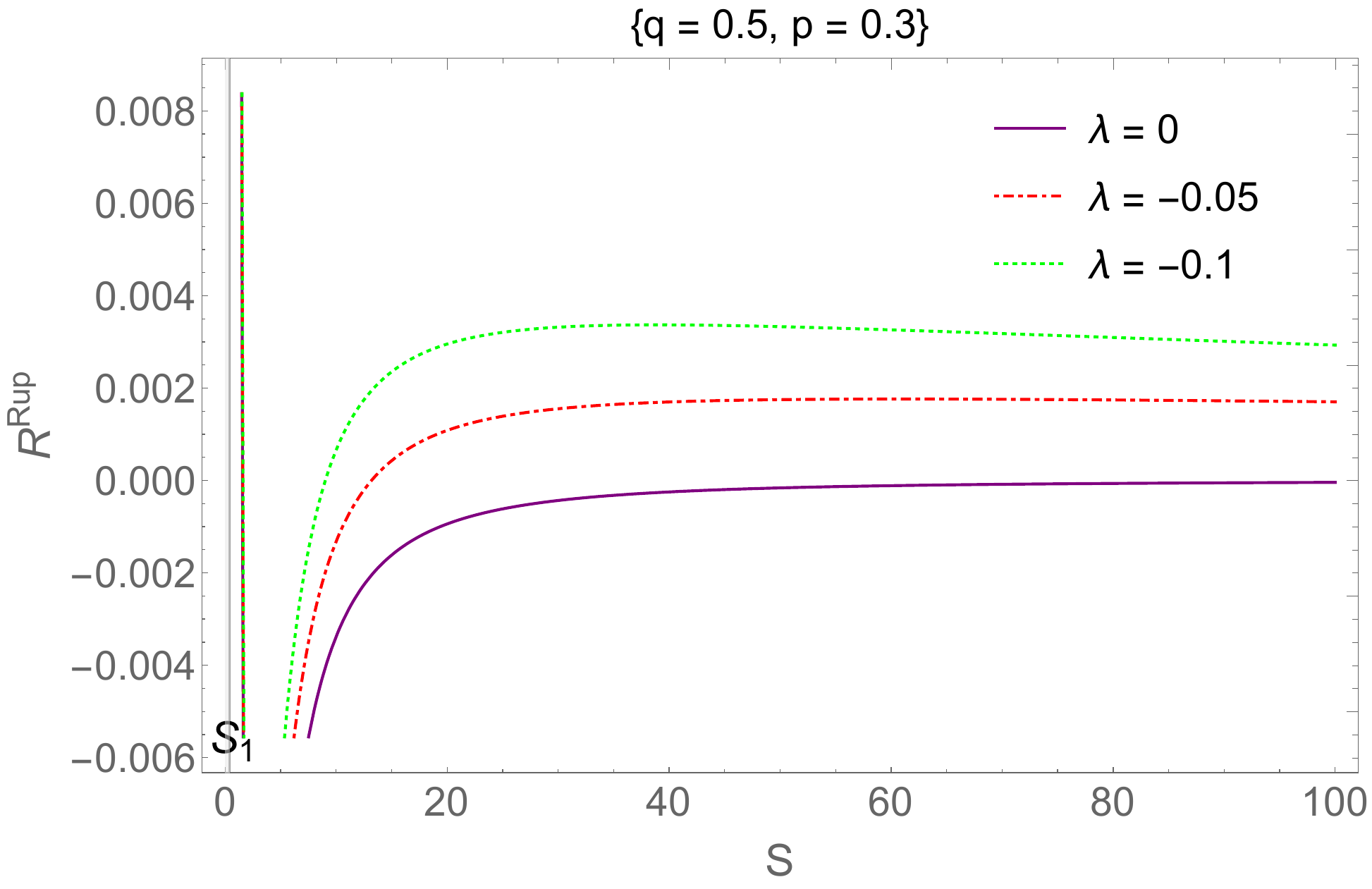} 
\caption{Ruppeiner curvature of the charged AdS BH as a function of entropy, for $\lambda > 0$ (upper panel) and $\lambda<0$ (lower panel). The shaded regions correspond to the unphysical regime where $T<0$.
}
\label{RRup3}
\end{center}
\end{figure}

\begin{figure}[t]
\begin{center}
\includegraphics[width=8cm]{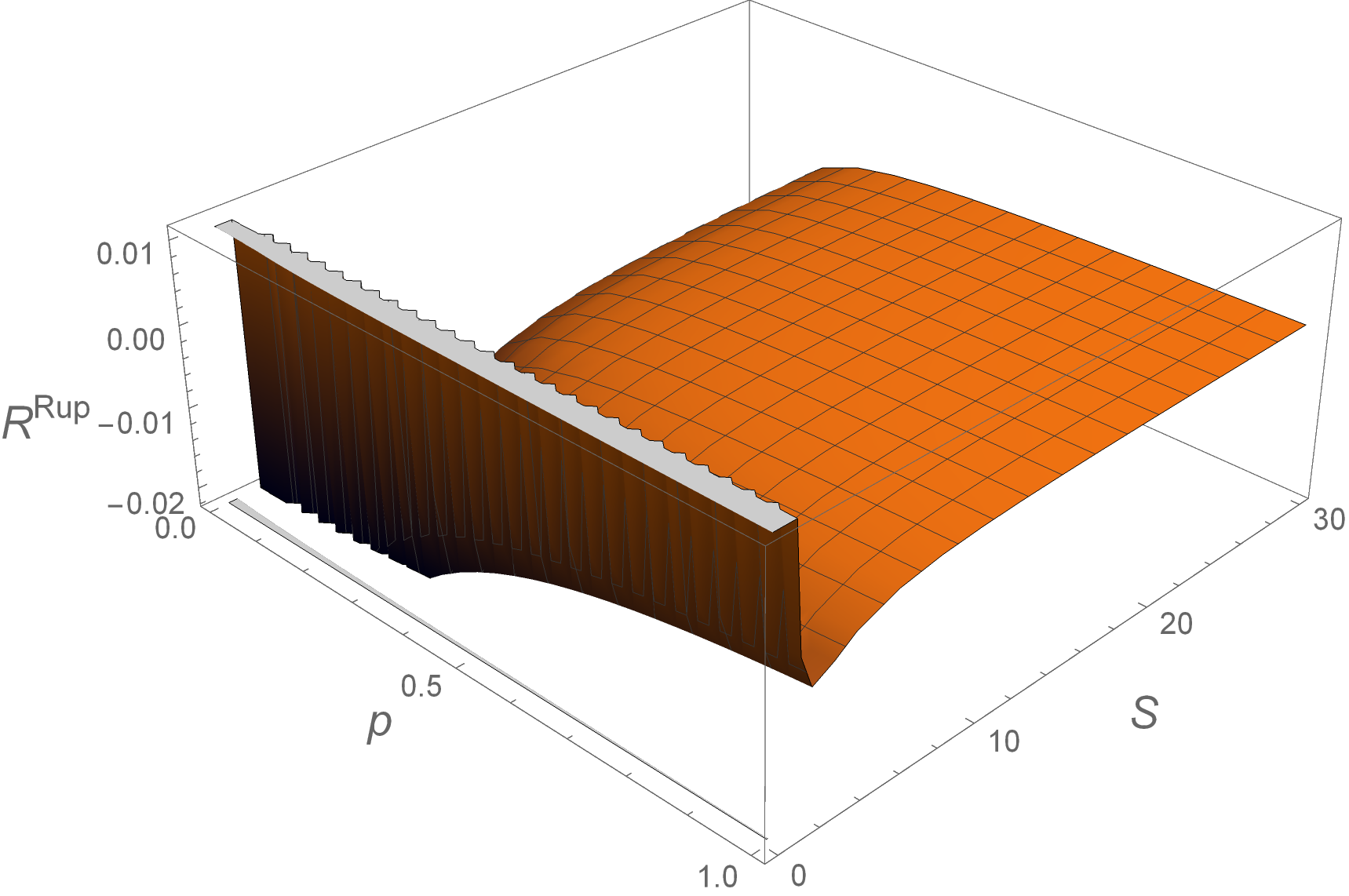} 

\vspace{0.5cm}
\includegraphics[width=8cm]{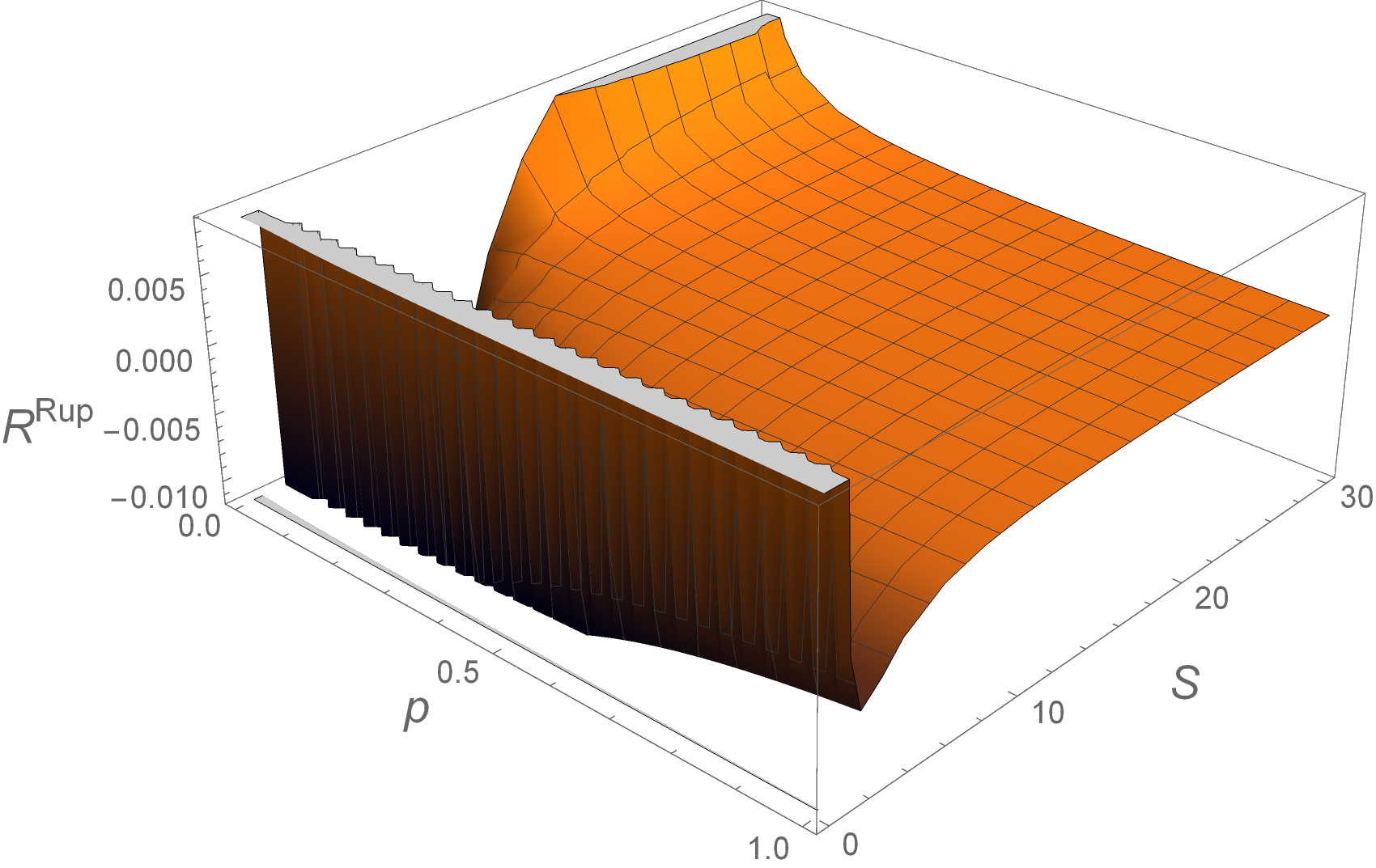} 
\caption{3D plot of Ruppeiner curvature of charged AdS BH as a function of entropy and pressure, for fixed $q=0.5$,  $\lambda = 0.05$ (upper panel) and $\lambda=-0.05$ (lower panel).
}
\label{3D3}
\end{center}
\end{figure}

We conclude our analysis by extending the previous study to include different values of \( p \) and \( q \). The corresponding results are displayed in Fig.~\ref{RRup3} and Fig.~\ref{3D3}, which reveal qualitatively similar features and exhibit a pattern analogous to that observed in Fig.~\ref{RRup2} and Fig.~\ref{3D2}, respectively.

\section{Conclusions and Outlook}
\label{Conc}

Conformal Killing Gravity is a  modification of GR, which encapsulates the nontrivial contributions from spacetime conformal curvature. In this work, we analyzed the impact of CKG on BH thermodynamics.  Specifically, to examine the stability and phase structure of Schwarzschild-AdS and charged AdS BHs within this framework, we studied the behavior of key thermodynamic quantities, including the mass, temperature, heat capacity and Gibbs free energy. We showed that, depending on the value and sign of the parameter \( \lambda \), different BH configurations can emerge, with the corrective effects becoming increasingly relevant at large distance scales, where the modified gravity term dominate the gravitational dynamics. 

Interestingly, we found that for \( \lambda > 0 \) and sufficiently large values of entropy (or, equivalently, large BH radius), the BH mass exhibits a peculiar decreasing trend as a function of entropy. This behavior signals a potentially pathological regime, as it corresponds to negative values of the absolute temperature associated with Hawking radiation. Such a feature may indicate the breakdown of the classical description, highlighting the necessity of adopting a more complete theoretical framework in order to accurately capture the thermodynamic behavior of the system beyond this regime.

Nontrivial effects also arise at the level of thermodynamic stability, both local and global, with the parameter \( \lambda \) governing the divergences in the heat capacity as well as the sign changes in both the heat capacity and the Gibbs free energy. These behaviors are key indicators of phase transitions and shifts in stability. Indeed, divergences in the heat capacity signal the presence of second-order phase transitions, associated with changes in the local thermodynamic stability of the BH.
The structure of these thermodynamic quantities suggests the presence of Hawking-Page-like transitions, where the system undergoes a transition between thermal AdS space and a BH phase depending on the temperature (or entropy) and the parameters \( \lambda \), \( p \), and \( q \). Notably, the inclusion of the modified gravity term modifies the location and nature of these transitions. 

Finally, adopting the perspective of BHs as thermodynamic systems endowed with an intrinsic microscopic structure, we employed the framework of geometrothermodynamics to investigate the interactions among their underlying constituents. To this end, we computed the Ruppeiner scalar curvature, whose sign provides insight into the character of these interactions. Once again, we demonstrated that, depending on the sign of \( \lambda \), one can distinguish regimes in which \( R^{\mathrm{Rup}} \) is positive, negative or vanishes. 
The transition between these regimes  provides valuable insight into the evolution of microscopic correlations and the underlying statistical structure of BH thermodynamics within the CKG framework. In particular, the emergence of a repulsive phase for $\lambda<0$, which is absent in standard Schwarzschild-AdS BHs, highlights the significant role of the modified gravity correction in modifying the effective microstructure of the system.

Further aspects remain to be explored. In particular, it would be of considerable interest to investigate a potential connection between the large-scale modifications to BH thermodynamics induced by CKG, arising from purely geometric and gravitational considerations, and the effects associated with non-extensive Tsallis entropy~\cite{Tsallis,Luciano:2023fyr,Dago}, which has been shown to emerge naturally in self-gravitating systems characterized by long-range interactions and fractal phase space structures.

Additionally, these investigations may be complemented by insights from the Extended Uncertainty Principle (EUP), a large-scale modification of the standard Heisenberg relation~\cite{EUP1,EUP2,Gine:2020izd}. In fact, the EUP implies a minimum measurable momentum and becomes relevant at cosmological or infrared scales, aligning naturally with the regime in which modified gravity corrections become significant. Intriguingly, EUP-based corrections have been shown to modify BH thermodynamics by introducing logarithmic or power-law deviations in entropy and temperature~\cite{BHEUP1,BHEUP2}. Establishing links between EUP-induced corrections, non-extensive entropy and CKG may offer a unified framework for describing gravitational systems beyond the classical regime, with implications for infrared quantum gravity and the thermodynamics of large-scale structures. A detailed analysis is underway and will be presented elsewhere.

\acknowledgments 
The authors express their sincere gratitude to Carlo Alberto Mantica and Bayram Tekin for their valuable comments and suggestions on the original version of the manuscript.
The research of GGL is supported by the postdoctoral fellowship program of the University of Lleida. GGL gratefully acknowledges the contribution of the LISA Cosmology Working Group (CosWG), as well as support from the COST Actions CA21136 - \textit{Addressing observational tensions in cosmology with systematics and fundamental physics (CosmoVerse)} - CA23130, \textit{Bridging high and low energies in search of quantum gravity (BridgeQG)} and CA21106 -  \textit{COSMIC WISPers in the Dark Universe: Theory, astrophysics and experiments (CosmicWISPers)}.

\end{document}